\documentclass[12pt]{article}
\usepackage{amsmath,amssymb,epsfig,ulem,feynarts}
\usepackage[bookmarks=false]{hyperref}
\usepackage{psfrag}
\allowdisplaybreaks  

\def\bm#1{\mbox{\boldmath$#1$\unboldmath}} 

\def\Mkk{M_{\rm KK}} 
\newcommand{\ord}{{\cal O}} 
\newcommand{\ie}{{\it i.e.}}

\newcommand{\beq}{\begin{equation}}
\newcommand{\eeq}{\end{equation}}
\newcommand{\bea}{\begin{eqnarray}} 
\newcommand{\eea}{\end{eqnarray}}
\newcommand{\Fig}[1]{Figure~\ref{#1}}

\newcommand{\Tab}[1]{Table~\ref{#1}}
\newcommand{\Sec}[1]{Section~\ref{#1}}
\newcommand{\App}[1]{Appendix~\ref{#1}}
\newcommand{\eq}[1]{(\ref{#1})} 
\newcommand{\ff}{f\hspace{-0.4em}f}
\newcommand{\CF}{C_F} 
\newcommand{\TF}{T_F} 

\newcommand{\Nc}{N_c} 
\newcommand{\GeV}{\hspace{1mm} {\rm GeV}}
\newcommand{\TeV}{\hspace{1mm} {\rm TeV}} 
\newcommand{\AFBt}{A_{\rm FB}^{t}}
\newcommand{\Atc}{A_c^{t}}
\newcommand{\sigtot}{\sigma_{t\bar{t}}}
\newcommand{\dsig}{d\sigma_{t\bar{t}}/dM_{t\bar{t}}}
\newcommand{\Mtt}{M_{t\bar{t}}}

\begin{document}

\begin{titlepage}

\begin{flushright}
MZ-TH/10-28\\
\end{flushright}

\vspace{15pt}
\begin{center}
  \Large\bf Top-Quark Forward-Backward Asymmetry in Randall-Sundrum
  Models Beyond the Leading Order
\end{center}

\vspace{5pt}
\begin{center}
{\sc 
M.~Bauer, F.~Goertz, U.~Haisch, T.~Pfoh and S.~Westhoff}\\
\vspace{10pt} {\sl
Institut f\"ur Physik (THEP),
Johannes Gutenberg-Universit\"at \\
D-55099 Mainz, Germany}
\end{center}

\vspace{10pt}
\begin{abstract}\vspace{2pt} \noindent 
  We calculate the $t \bar t$ forward-backward asymmetry, $\AFBt$, in
  Randall-Sundrum (RS) models taking into account the dominant
  next-to-leading order (NLO) corrections in QCD. At Born level we
  include the exchange of Kaluza-Klein (KK) gluons and photons, the
  $Z$ boson and its KK excitations, as well as the Higgs boson,
  whereas beyond the leading order (LO) we consider the interference
  of tree-level KK-gluon exchange with one-loop QCD box diagrams and
  the corresponding bremsstrahlungs corrections. We find that the
  strong suppression of LO effects, that arises due to the elementary
  nature and the mostly vector-like couplings of light quarks, is
  lifted at NLO after paying the price of an additional factor of
  $\alpha_s/(4 \pi)$. In spite of this enhancement, the resulting RS
  corrections in $\AFBt$ remain marginal, leaving the predicted
  asymmetry SM-like. As our arguments are solely based on the
  smallness of the axial-vector couplings of light quarks to the
  strong sector, our findings are model-independent and apply to many
  scenarios of new physics that address the flavor problem via
  geometrical sequestering.
\end{abstract}

\vfill
\end{titlepage}

\tableofcontents

\section{Introduction}
\label{sec:intro}

The top quark is the heaviest particle in the Standard Model (SM) of
particle physics. Its large mass suggests that it might be deeply
connected to the mechanism driving electroweak symmetry breaking.
Detailed experimental studies of the top-quark properties are thus
likely to play a key role in unravelling the origin of mass, making
top-quark observables one of the cornerstones of the Fermilab Tevatron
and CERN Large Hadron Collider (LHC) physics programmes.

Up to now, the CDF and D{\O} experiments at the Tevatron have
collected thousands of top-quark pair events, which allowed them to
measure the top-quark mass, $m_t$, and its total inclusive cross
section, $\sigtot$, with an accuracy of below $1\%$ \cite{top:2009ec}
and $10\%$ \cite{CDFnotetot, D0notetot}, respectively. While these
measurements are important in their own right, from the point of view
of searches for physics beyond the SM, determinations of kinematic
distributions and charge asymmetries in $t \bar t$ production are more
interesting, since these observables are particularly sensitive to
non-standard dynamics. Such searches have been performed at the
Tevatron \cite{Aaltonen:2007dz, Aaltonen:2007dia, Abazov:2008ny}, and
a result for the $t \bar t$ invariant mass spectrum, $\dsig$, has been
recently obtained from data collected at CDF \cite{Bridgeman:2008zz,
  Aaltonen:2009iz}. The forward-backward asymmetry, $\AFBt$, has also
been measured \cite{Schwarz:2006ud, Abazov:2007qb, Aaltonen:2008hc,
  publicCDF, CDFbrandnew} and constantly found to be larger than
expected. In the laboratory ($p \bar p$) frame the most recent CDF
result reads
\beq \label{eq:AFBexp}
\left ( \AFBt \right )_{\rm exp}^{p\bar p} = (15.0 \pm
5.0_{\rm{\hspace{0.5mm} stat.}} \pm 2.4_{\rm{\hspace{0.5mm} syst.}} )
\, \% \,,
\eeq
where the quoted uncertainties are of statistical and systematical
origin, respectively.\footnote{Very recently {D\O} reported a
  measurement of $(\AFBt)_{\rm exp}^{\rm obs.} = (8 \pm 4_{\rm stat.}
  \pm 1_{\rm syst.})\%$ for $t \bar t$ events that satisfy the
  experimental acceptance cuts \cite{D0brandnew}. The corresponding
  SM prediction reads $(\AFBt)_{\rm SM}^{\rm obs.} = \left
    (1^{+2}_{-1} \right )\%$ and is similarly below the observed
  value.}

At leading order (LO) in QCD, the charge-asymmetric cross section is
zero within the SM. Starting from ${\cal O} (\alpha_s^3)$ or
next-to-leading order (NLO) onward, the quantity $\AFBt$ receives
non-vanishing contributions. These arise from the interference of
tree-level gluon exchange with one-loop QCD box diagrams and the
interference of initial- and final-state radiation. Including NLO as
well as electroweak corrections \cite{Kuhn:1998jr, Kuhn:1998kw}, the
SM prediction in the $p \bar p$ frame for the inclusive asymmetry is
\cite{Antunano:2007da}
\beq \label{eq:AFBSM}
\left ( \AFBt \right )_{\rm SM}^{p \bar p} = (5.1 \pm 0.6) \, \%
\,,
\eeq
where the total error includes the individual uncertainties due to
different choices of the parton distribution functions (PDFs), the
factorization and renormalization scales, and a variation of $m_t$
within its experimental error. Recent theoretical determinations of
$(\AFBt)_{\rm SM}$, that include the resummation of logarithmically
enhanced threshold effects at NLO \cite{Almeida:2008ug} and
next-to-next-to-leading order (NNLO) \cite{Ahrens:2010zv}, are in
substantial agreement with the latter number. These results together
with general theoretical arguments \cite{Melnikov:2010iu} suggest
that the value (\ref{eq:AFBSM}) is robust with respect to higher-order
QCD corrections, making it a firm SM prediction. 

Although the discrepancy between the experimental (\ref{eq:AFBexp})
and the theoretical (\ref{eq:AFBSM}) value of $\AFBt$ is not
significant given the sizable statistical error,\footnote{With respect
  to the previously published CDF result \cite{publicCDF}, the
  updated measurement (\ref{eq:AFBexp}) is in better agreement with
  the SM prediction. The former $2 \sigma$ discrepancy is now a $1.7
  \sigma$ deviation.} the persistently large values of the observed
asymmetry have triggered a lot of activity in the theory community
\cite{Djouadi:2009nb, Ferrario:2009bz, Jung:2009jz, Cheung:2009ch,
  Frampton:2009rk, Shu:2009xf, Arhrib:2009hu, Dorsner:2009mq,
  Jung:2009pi, Cao:2009uz, Barger:2010mw, Cao:2010zb, Xiao:2010hm,
  Chivukula:2010fk}. Many scenarios beyond the SM impact $\AFBt$
already at LO by tree-level exchange of new heavy particles with
axial-vector couplings to fermions. However, it turns out to be
difficult in general to explain the large central experimental value,
since any viable model must simultaneously avoid giving rise to
unacceptably large deviations in $\sigtot$ and/or $\dsig$, which both
show no evidence of non-SM physics. The first class of proposed models
envisions new physics in the $t$ channel (or $u$ channel) with large
flavor-violating couplings induced either by vector-boson exchange,
namely $W^\prime$ \cite{Cheung:2009ch, Barger:2010mw, Cao:2010zb} and
$Z^\prime$ bosons \cite{Jung:2009jz, Jung:2009pi, Cao:2009uz,
  Barger:2010mw, Cao:2010zb, Xiao:2010hm}, or by exchange of color
singlet, triplet, or sextet scalars \cite{Shu:2009xf, Arhrib:2009hu,
  Dorsner:2009mq, Jung:2009pi,  Cao:2009uz, Cao:2010zb}. On general
grounds, it is not easy to imagine how the necessary flavor-changing
couplings can be generated naturally without invoking {\it ad hoc}
assumptions. A second class of models involves $s$-channel tree-level
exchange of new vector states \cite{Djouadi:2009nb, Ferrario:2009bz,
  Frampton:2009rk, Jung:2009pi, Cao:2009uz, Cao:2010zb,
  Chivukula:2010fk}, preferably color octets to maximize their
interference with QCD, that exhibit sizable axial-vector couplings to
both the light quarks, $g_A^q$, and the top quark, $g_A^t$. In order
to achieve a positive shift in $\AFBt$ the new vectors have to couple
to the first and the third generation of quarks with opposite
axial-vector couplings \cite{Ferrario:2008wm}, implying $g_A^q
\hspace{0.25mm} g_A^t < 0$. Examples of theories that were found to
lead to a positive shift in the charge asymmetry are scenarios with a
warped extra dimension \cite{Djouadi:2009nb} and flavor non-universal
chiral color models \cite{Frampton:2009rk, Chivukula:2010fk}, both
featuring heavy exotic partners of the SM gluon.

The purpose of this article is to show that, in the wide class of
scenarios beyond the SM that are dominated by virtual exchange of
vector bosons in the $s$ channel, the NLO corrections to $\AFBt$ can
exceed the LO corrections if the axial-vector couplings to the light
quarks are suppressed.\footnote{The importance of NLO corrections has
  been briefly mentioned in \cite{Ferrario:2009ee}, which discusses
  the charge asymmetry in the exclusive channel $p\bar p\rightarrow
  t\bar t X$.} We will argue that this observation applies in
particular to new-physics scenarios that explain the hierarchical
structures observed in the masses and mixing of the SM fermions
geometrically (which is in one-to-one correspondence to the
Froggatt-Nielsen mechanism \cite{Froggatt:1978nt}). Since this way of
generating fermion hierarchies also entails a suppression of harmful
flavor-changing neutral currents (FCNCs), sequestering flavor is
likely to be an integral part of any theory where the solution to the
fermion puzzle is associated to a new-physics scale low enough to be
directly testable at the LHC. While our considerations are for most of
the part general, we find it instructive to elucidate them by working
out in detail the relevant LO and NLO corrections to $\AFBt$ that
arise in Randall-Sundrum (RS) models \cite{Randall:1999ee}. This
class of constructions can be regarded as the prototype of
non-standard scenarios harnessing the idea of split fermions
\cite{ArkaniHamed:1999dc} by locating the left- and right-handed
fermions at different places in a warped extra dimension. While the
localization pattern gives rise to the necessary axial-vector
couplings $g_A^{q,t}$ of Kaluza-Klein (KK) gluons to the SM quarks,
the couplings $g_A^q$ turn out to be doubly suppressed: first, because
the light quarks reside in the ultraviolet (UV) and, second, because
their wave functions of different chiralities are localized nearby in
the fifth dimension. The light-quark vector couplings $g_V^q$ do not
suffer from the latter type of suppression. In contrast, the top-quark
axial-vector and vector couplings, $g_{A, V}^t$, can be sizable due to
the large overlap of the third-generation up-type quark wave functions
with the ones of the KK gluons, all of which are peaked in the
infrared (IR). Given the strong suppression of $g_A^q \hspace{0.25mm}
g_A^t$ in the RS framework, it is then natural to ask if the effects
in $\AFBt$ that depend on the product $g_V^q \hspace{0.25mm} g_V^t$ of
vector couplings are phenomenologically more important, despite the
fact that this combination enters the prediction for the charge
asymmetry first at the one-loop level. This question can only be
answered by studying the interplay of new-physics contributions to
$t\bar{t}$ production at LO and NLO in detail, and this is exactly
what we will do in the following.

This article is organized as follows. After reviewing in
\Sec{sec:prelim} the kinematics and the structure of the various $t
\bar t$ observables in the SM, we present in Section~\ref{sec:AFBtRS}
the calculation of the RS corrections to the total cross section
$\sigtot$ and the forward-backward asymmetry $\AFBt$. At LO we compute
the tree-level exchange of KK gluons and photons, the $Z$ boson and
its KK resonances, as well as the Higgs boson, while at NLO we take
into account the interference of the dominant tree-level KK-gluon
exchange with the one-loop QCD box graphs supplemented by real gluon
emission.  In order to keep our discussion as general as possible, we
perform the calculation in an effective-field theory (EFT) obtained
after integrating out the heavy KK states. As a result, our analytic
formulas are applicable to a wide class of scenarios with non-standard
dynamics above the electroweak scale. We then discuss the structure of
the LO and NLO corrections that arise from the exchange of KK gluons
in the $s$ channel. As anticipated, we find that, due to the strong
suppression of the axial-vector couplings of light quarks, the ${\cal
  O} (\alpha^3_s)$ corrections to $\AFBt$ typically dominate over the
${\cal O} (\alpha_s^2)$ contributions in warped models. Our detailed
numerical analysis of Section~\ref{sec:numerics} confirms this general
finding, but also shows that RS effects are too small to explain the
anomalously large value of the $t \bar t$ charge asymmetry measured at
the Tevatron. We conclude in Section~\ref{sec:concl}. In a series of
appendices we collect details on the phase-space factors appearing in
the Higgs-boson contribution, give the analytic expressions for the
relevant Wilson coefficients, present compact formulas for the
renormalization group (RG) evolution of the relevant Wilson
coefficients, and detail the parameter points used in our numerical
analysis.

\section{Top-Antitop Production in the SM}
\label{sec:prelim}

At the Tevatron $t\bar{t}$ pairs are produced in collisions of protons
and antiprotons, $p\bar{p}\to t\bar{t}X$. Within the SM the hadronic
process receives partonic Born-level contributions from
quark-antiquark annihilation and gluon fusion
\beq \label{eq:SMtreeprocesses}
\begin{split}
  q(p_1) + \bar{q}(p_2) &\to t(p_3) + \bar{t}(p_4) \,, \\
  g(p_1) + g(p_2) &\to t(p_3) + \bar{t}(p_4) \,,
\end{split}
\eeq
where the four-momenta $p_{1,2}$ of the initial state partons can be
expressed as the fractions $x_{1,2}$ of the four-momenta $P_{1,2}$ of
the colliding hadrons, $p_{1,2}=x_{1,2} \hspace{0.25mm} P_{1,2}$, and
$s = (P_1+P_2)^2$ denotes the hadronic center-of-mass (CM) energy
squared. The partonic cross section is a function of the kinematic
invariants
\beq \label{eq:invariants}
  \hat{s} = (p_1 + p_2)^2 \,, 
  \qquad t_1 = (p_1 - p_3)^2 - m_t^2 \,,
  \qquad u_1 = (p_2 - p_3)^2 - m_t^2 \,,
\eeq
and momentum conservation at Born level implies that $\hat{s} + t_1 +
u_1 = 0$.

Since we will be interested in the differential cross section with
respect to the invariant mass $\Mtt = \sqrt{(p_3 + p_4)^2}$ of the $t
\bar t$ pair and the angle $\theta$ between $\vec{p}_1$ and
$\vec{p}_3$ in the partonic CM frame, we express $t_1$ and $u_1$ in
terms of $\theta$ and the top-quark velocity $\beta$,
\beq \label{eq:t1u1} 
  t_1 = -\frac{\hat s}{2} ( 1 - \beta \cos\theta ) \,, \qquad 
  u_1 = -\frac{\hat s}{2} ( 1 + \beta \cos\theta ) \,, \qquad 
  \beta = \sqrt{1-\rho} \,, \qquad \rho = \frac{4m_t^2}{\hat s} \,.
\eeq
The hadronic differential cross section may then be written as
\beq \label{eq:dsdc}
\frac{d\sigma^{p\bar{p}\rightarrow t\bar t X}}{d\cos\theta} =
\frac{\alpha_s}{m_t^2} \sum_{i,j} \int_{4m_t^2}^s \frac{d \hat s}{s}
\, \ff_{ij}\big(\hat s/s,\mu_f\big) \, K_{ij} \left (
  \frac{4m_t^2}{\hat s},\cos\theta,\mu_f \right ) \,,
\eeq
where $\mu_f$ denotes the factorization scale and we have introduced
the parton luminosity functions
\beq \label{eq:luminosities}
\ff_{ij}(y,\mu_f) = \int_y^1 \frac{dx}{x} \, f_{i/p}(x,\mu_f) \,
f_{j/\bar{p}}(y/x,\mu_f) \,.
\eeq
The luminosities for $ij = q \bar q, \bar q q$ are understood to be
summed over all species of light quarks, and the functions
$f_{i/p}(x,\mu_f)$ ($f_{i/\bar{p}}(x,\mu_f)$) are the universal
non-perturbative PDFs, which describe the probability of finding the
parton $i$ in the proton (antiproton) with longitudinal momentum
fraction $x$. The hard-scattering kernels $K_{ij}
(\rho,\cos\theta,\mu_f)$ are related to the partonic cross sections
and have a perturbative expansion in $\alpha_s$ of the form
\beq \label{eq:Cijexp} 
K_{ij} (\rho,\cos\theta,\mu_f) = \sum_{n = 0}^\infty \left (
  \frac{\alpha_s}{4\pi} \right )^n \, K_{ij}^{(n)}
(\rho,\cos\theta,\mu_f) \,.
\eeq
In the SM only the hard-scattering kernels with $ij = q \bar q, \bar q
q, gg$ are non-zero at LO in $\alpha_s$. By calculating the amplitudes
corresponding to $s$-channel gluon exchange one finds
\bea \label{eq:SMLO}
\begin{split}
  K_{q\bar{q}}^{(0)} &= \alpha_s\,\frac{\pi \beta
    \rho}{8}\,\frac{\CF}{\Nc}\,\left( \frac{t_1^2+u_1^2}{\hat{s}^2} +
    \frac{2m_t^2}{\hat s} \right) , \\[1mm]
K_{gg}^{(0)} &= \alpha_s\,\frac{\pi \beta \rho}{8(\Nc^2-1)} \left( \CF
  \, \frac{\hat{s}^2}{t_1u_1} - \Nc \right) \left[
  \frac{t_1^2+u_1^2}{\hat{s}^2} + \frac{4m_t^2}{\hat{s}} -
  \frac{4m_t^4}{t_1u_1} \right] ,
\end{split}
\eea 
and the coefficient $K_{\bar{q} q}^{(0)}$ is obtained from $K_{q
  \bar{q}}^{(0)}$ by replacing $\cos\theta$ with $-\cos\theta$. The
factors $\Nc = 3$ and $\CF = 4/3$ are the usual color
factors of $SU(3)_c$.

In the context of our work it will be convenient to follow
\cite{Almeida:2008ug} and to introduce charge-asymmetric ($a$) and
-symmetric ($s$) averaged differential cross sections. In the former
case, we define
\beq \label{eq:symasym}
  \frac{d \sigma_{a}}{d \cos{\theta}} \equiv \frac{1}{2}
  \left[\frac{d \sigma^{p\bar{p} \to t\bar{t} X}}{d \cos{\theta}}
 - \frac{d \sigma^{p\bar{p} \to \bar{t} t X}}{d \cos{\theta}}
  \right] \,, 
\eeq
with $d \sigma^{p\bar{p} \to t\bar{t} X}/d \cos{\theta}$ given in
\eq{eq:dsdc}. The corresponding expression for the charge-symmetric
averaged differential cross section $d\sigma_s/d\cos\theta$ is simply
obtained from the above by changing the minus into a plus sign. The
notation indicates that in the process labelled by the superscript
$p\bar{p} \to t\bar{t} X$ ($p\bar{p} \to \bar{t} t X$) the angle
$\theta$ corresponds to the scattering angle of the top (antitop)
quark in the partonic CM frame.  Using \eq{eq:symasym} one can derive
various physical observables in $t\bar{t}$ production. For example,
the total hadronic cross section is given by
\beq \label{eq:cs}
\sigtot =\int_{-1}^1 d\cos\theta \, \frac{d\sigma_s}{d\cos\theta} \,.
\eeq
We will mainly be interested in the total $t\bar{t}$ charge asymmetry
defined by
\beq \label{eq:chargeas}
A^t_c \equiv \frac{ \displaystyle \int_0^1 d \cos \theta \, \frac{d
    \sigma_a}{d \cos{\theta}}} { \displaystyle \int_0^1 d \cos \theta
  \, \frac{d\sigma_s}{d \cos{\theta}}} \,.
\eeq
Since QCD is symmetric under charge conjugation, which implies that
\beq \label{eq:charge}
\left. \frac{d \sigma^{p\bar{p} \to \bar{t} t X}}{d \cos{\theta}}
\right|_{\cos \theta = c} = \left. \frac{d \sigma^{p\bar{p} \to t
      \bar{t} X}}{d \cos{\theta}} \right|_{\cos \theta = - c} \,,
\eeq
for any fixed value $c$, the charge asymmetry can also be understood
as a forward-backward asymmetry
\beq\label{eq:fbas}
\Atc = \AFBt \equiv \frac{ \displaystyle \int_0^1 d \cos
  \theta \; \frac{d \sigma^{p\bar{p} \to t \bar{t} X}}{d \cos{\theta}}
  - \int_{-1}^0 d \cos \theta \; \frac{d \sigma^{p\bar{p} \to t
      \bar{t} X}}{d \cos{\theta}}} {\displaystyle \int_0^1 d \cos
  \theta \; \frac{d\sigma^{p\bar{p} \to t \bar{t} X}} {d \cos{\theta}}
  + \int_{-1}^0 d \cos \theta \; \frac{d\sigma^{p\bar{p} \to t\bar{t}
      X}}{d \cos{\theta}}} = \frac{\sigma_a}{\sigma_s}\, .
\eeq
For later convenience we express the asymmetric contribution to the
cross section as 
\beq \label{eq:sigmatotRSLO}
\sigma_a = \frac{\alpha_s}{m_t^2} \, \sum_{i,j} \int_{4m_t^2}^s
\frac{d \hat s}{s} \, \ff_{ij}\big( \hat s/s,\mu_f\big) \,
A_{ij} \left ( \frac{4m_t^2}{\hat s} \right ) \,.
\eeq
An analogous expression holds in the case of the symmetric
contribution $\sigma_s$ with the hard-scattering charge-asymmetric
coefficient $A_{ij} (4 m_t^2/\hat s)$ replaced by its symmetric
counterpart $S_{ij} (4 m_t^2/\hat s)$.

In the SM the LO coefficients of the symmetric part read
\beq \label{eq:SMsymLO}
\begin{split}
  S^{(0)}_{q\bar{q}} &= \alpha_s \, \frac{\pi\beta\rho}{27}\,(2+\rho)\,,\\
  S^{(0)}_{gg} &= \alpha_s\, \frac{\pi\beta\rho}{192} \left [
    \frac{1}{\beta} \, \ln \left ( \frac{1+\beta}{1-\beta} \right )
    \left ( 16 + 16 \rho + \rho^2\right ) - 28 -31\rho \right ] ,
\end{split}
\eeq
while the asymmetric contributions $A_{q\bar{q}}^{(0)}$ and
$A_{gg}^{(0)}$ both vanish identically. As we will explain in detail
in \Sec{sec:NLORS}, at NLO a non-zero coefficient $A_{q \bar q}^{(1)}$
is generated in the SM, which leads to a forward-backward asymmetry
that is suppressed by $\alpha_s/(4\pi)$ with respect to the symmetric
cross section.

\section{Cross Section and Asymmetry in RS Models}
\label{sec:AFBtRS}

The RS framework was originally proposed to explain the large
hierarchy between the electroweak and the Planck scales via
red-shifting in a warped fifth dimension. If the SM fermions and gauge
bosons are allowed to propagate in the bulk of the extra dimension
\cite{Davoudiasl:1999tf, Pomarol:1999ad, Grossman:1999ra,
  Chang:1999nh, Gherghetta:2000qt}, the RS model is in addition a
promising theory of flavor \cite{ArkaniHamed:1999dc, Grossman:1999ra,
  Gherghetta:2000qt, Huber:2000ie, Huber:2003tu,
  Agashe:2004cp}.\footnote{A list of further relevant references can
  be found in \cite{Bauer:2009cf}.}  Since the fermion zero modes are
exponentially localized either in the UV (light SM fermions) or IR
(heavy SM fermions), the effective Yukawa couplings resulting from
their wave-function overlap with the Higgs boson naturally exhibit
exponential hierarchies. In this way one obtains an extra-dimensional
realization \cite{Casagrande:2008hr, Blanke:2008zb} of the
Froggatt-Nielsen mechanism \cite{Froggatt:1978nt}, in which the flavor
structure is accounted for apart from ${\cal O}(1)$ factors. Another
important feature that follows from the structure of the overlap
integrals is that the effective coupling strength of KK gluons to
heavy quarks is enhanced relative to the SM couplings by a factor
$\sqrt{L}\,$ \cite{Davoudiasl:1999tf, Pomarol:1999ad} because the
involved fields are all localized in the IR. Here $L \equiv \ln \left
  (M_{\rm Pl}/M_W \right ) \approx \ln \left (10^{16} \right ) \approx
37$ denotes the logarithm of the warp factor, which is fixed by the
hierarchy between the electroweak ($M_W$) and the fundamental Planck
($M_{\rm Pl}$) scales. Since left- and right-handed fermions are
localized at different points in the bulk, the KK-gluon couplings to
quarks are in general not purely vector-like, but receive
non-vanishing axial-vector components. These couplings generate a
charge asymmetry in top-quark pair production at LO, which is
associated to quark-antiquark annihilation $q \bar q \to t \bar t$ and
proceeds through tree-level exchange of KK gluons in the $s$
channel. In the RS model, further corrections to $\AFBt$ arise from
the fact that the couplings of KK gluons and photons, the $Z$ boson
and its KK excitations, as well as the Higgs boson are flavor
non-diagonal, leading to the flavor-changing $u \bar u \to t \bar t$
transition which affects the $t$ channel.\footnote{In principle, also
  the $d \bar d \to t \bar t$ transition receives corrections due to
  the $t$-channel exchange of the $W$ boson and its KK partners. We
  have explicitly verified that these effects are negligibly small for
  viable values of $\Mkk$.  Therefore we will ignore them in the
  following.} The corresponding diagrams are shown in
\Fig{fig:RSBorn}. On the other hand, the gluon-fusion channel $g g \to
t \bar t$ does not receive a correction at Born level, since owing to
the orthonormality of gauge-boson wave functions the coupling of two
gluons to a KK gluon is zero.

\subsection{Calculation of LO Effects}
\label{sec:LORS}

Since the KK scale $\Mkk$ is at least of the order of a few times the
electroweak scale, virtual effects appearing in RS models can be
decribed by means of an effective low-energy theory consisting out of
dimension-six operators. In the case at hand, the effective Lagrangian
needed to account for the effects of intermediate vector and scalar
states reads
\beq \label{eq:Leff}
{\cal L}_{\rm eff} = \sum_{q,u} \sum_{A, B = L, R} \Big [
\hspace{0.25mm} C_{q \bar q, AB}^{(V, 8)} \hspace{0.25mm} Q_{q \bar q,
  AB}^{(V, 8)} + C_{t \bar u, AB}^{(V, 8)} \hspace{0.25mm} Q_{t
  \bar u, AB}^{(V, 8)}  + C_{t \bar u, AB}^{(V, 1)} \hspace{0.25mm} Q_{t \bar
  u, AB}^{(V, 1)} + C_{t \bar u, AB}^{(S, 1)} \hspace{0.25mm}
Q_{t \bar u, AB}^{(S, 1)} \hspace{0.25mm} \Big ] \,,
\eeq
where 
\beq \label{eq:operators}
\begin{split}
  Q_{q \bar q, AB}^{(V, 8)} & = (\bar q \hspace{0.25mm} \gamma_\mu
  \hspace{0.25mm} T^a \hspace{0.25mm} P_A \hspace{0.25mm} q) (\bar t
  \hspace{0.25mm} \gamma^\mu \hspace{0.25mm} T^a \hspace{0.25mm}
  P_B \hspace{0.25mm} t) \,, \\
  Q_{t \bar u, AB}^{(V, 8)} & = (\bar u \hspace{0.25mm} \gamma_\mu
  \hspace{0.25mm} T^a \hspace{0.25mm} P_A \hspace{0.25mm} t) (\bar t
  \hspace{0.25mm} \gamma^\mu \hspace{0.25mm} T^a \hspace{0.25mm} P_B
  \hspace{0.25mm} u) \,, \\
   Q_{t \bar u, AB}^{(V, 1)} & = (\bar u \hspace{0.25mm} \gamma_\mu
  \hspace{0.25mm} P_A \hspace{0.25mm} t) (\bar t \hspace{0.25mm}
  \gamma^\mu \hspace{0.25mm} P_B \hspace{0.25mm} u) \,, \\
  Q_{t \bar u, AB}^{(S, 1)} & = (\bar u \hspace{0.25mm} P_A
  \hspace{0.25mm} t) (\bar t \hspace{0.25mm} P_B \hspace{0.25mm} u)
  \,,
\end{split} 
\eeq 
and the sum over $q$ ($u$) involves all light (up-type) quark
flavors. In addition, $P_{L,R} = (1 \mp \gamma_5)/2$ project onto
left- and right-handed chiral quark fields, and $T^a$ are the
generators of $SU(3)_c$ normalized such that ${\rm Tr} \left ( T^a T^b
\right ) = \TF \, \delta_{ab}$ with $\TF = 1/2$. The superscripts $V$
and $S$ ($8$ and $1$) label vector and scalar (color-octet
and -singlet) contributions, respectively.

\begin{figure}
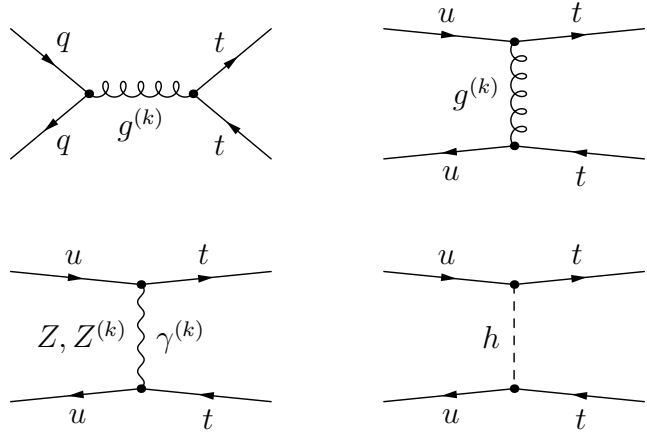

\begin{center}
\vspace{-8mm}
\mbox{
\unitlength=0.25bp
\begin{feynartspicture}(432,504)(1,1)
\FADiagram{}
\FAProp(0.,15.)(6.,10.)(0.,){/Straight}{1}
\FALabel(3.51229,13.2107)[bl]{$q$}
\FAProp(0.,5.)(6.,10.)(0.,){/Straight}{-1}
\FALabel(3.51229,6.78926)[tl]{$q$}
\FAProp(20.,15.)(14.,10.)(0.,){/Straight}{-1}
\FALabel(16.4877,13.2107)[br]{$t$}
\FAProp(20.,5.)(14.,10.)(0.,){/Straight}{1}
\FALabel(16.4877,6.78926)[tr]{$t$}
\FAProp(6.,10.)(14.,10.)(0.,){/Cycles}{0}
\FALabel(10.,8.93)[t]{$g^{(k)}$}
\FAVert(6.,10.){0}
\FAVert(14.,10.){0}
\end{feynartspicture}
}
\hspace{6mm}
\mbox{
\unitlength=0.25bp
\begin{feynartspicture}(432,504)(1,1)
\FADiagram{}
\FAProp(0.,15.)(10.,14.)(0.,){/Straight}{1}
\FALabel(4.84577,15.5623)[b]{$u$}
\FAProp(0.,5.)(10.,6.)(0.,){/Straight}{-1}
\FALabel(5.15423,4.43769)[t]{$u$}
\FAProp(20.,15.)(10.,14.)(0.,){/Straight}{-1}
\FALabel(14.8458,15.5623)[b]{$t$}
\FAProp(20.,5.)(10.,6.)(0.,){/Straight}{1}
\FALabel(15.1542,4.43769)[t]{$t$}
\FAProp(10.,14.)(10.,6.)(0.,){/Cycles}{0}
\FALabel(8.93,10.)[r]{$g^{(k)}$}
\FAVert(10.,14.){0}
\FAVert(10.,6.){0}
\end{feynartspicture}
}

\vspace{-1.25cm}

\mbox{
\unitlength=0.25bp
\begin{feynartspicture}(432,504)(1,1)
\FADiagram{}
\FAProp(0.,15.)(10.,14.)(0.,){/Straight}{1}
\FALabel(4.84577,15.5623)[b]{$u$}
\FAProp(0.,5.)(10.,6.)(0.,){/Straight}{-1}
\FALabel(5.15423,4.43769)[t]{$u$}
\FAProp(20.,15.)(10.,14.)(0.,){/Straight}{-1}
\FALabel(14.8458,15.5623)[b]{$t$}
\FAProp(20.,5.)(10.,6.)(0.,){/Straight}{1}
\FALabel(15.1542,4.43769)[t]{$t$}
\FAProp(10.,14.)(10.,6.)(0.,){/Sine}{0}
\FALabel(8.93,10.)[r]{$Z, Z^{(k)}$}
\FALabel(14.93,10.)[r]{$\gamma^{(k)}$}
\FAVert(10.,14.){0}
\FAVert(10.,6.){0}
\end{feynartspicture}
}
\hspace{6mm}
\mbox{
\unitlength=0.25bp
\begin{feynartspicture}(432,504)(1,1)
\FADiagram{}
\FAProp(0.,15.)(10.,14.)(0.,){/Straight}{1}
\FALabel(4.84577,15.5623)[b]{$u$}
\FAProp(0.,5.)(10.,6.)(0.,){/Straight}{-1}
\FALabel(5.15423,4.43769)[t]{$u$}
\FAProp(20.,15.)(10.,14.)(0.,){/Straight}{-1}
\FALabel(14.8458,15.5623)[b]{$t$}
\FAProp(20.,5.)(10.,6.)(0.,){/Straight}{1}
\FALabel(15.1542,4.43769)[t]{$t$}
\FAProp(10.,14.)(10.,6.)(0.,){/ScalarDash}{0}
\FALabel(8.93,10.)[r]{$h$}
\FAVert(10.,14.){0}
\FAVert(10.,6.){0}
\end{feynartspicture}
}
\end{center}
\vspace{-1.5cm}
\begin{center}
  \parbox{15.5cm}{\caption{\label{fig:RSBorn} Upper row: Tree-level
      contributions to the $q \bar q \to t \bar t$ (left) and the $u
      \bar u \to t \bar t$ (right) transition arising from $s$- and
      $t$-channel exchange of KK gluons. Lower row: Tree-level
      contributions to the $u \bar u \to t \bar t$ transition arising
      from $t$-channel exchange of the $Z$ boson, of KK photons and
      $Z$ bosons as well as of the Higgs boson. The $s$-channel
      ($t$-channel) amplitudes receive corrections from all light up-
      and down-type (up-type) quark flavors.}}
\end{center}
\end{figure}

Using the effective Lagrangian \eq{eq:Leff} it is straightforward to
calculate the interference between the tree-level matrix element
describing $s$-channel SM gluon exchange and the $s$- and $t$-channel
new-physics contributions arising from the Feynman graphs displayed in
\Fig{fig:RSBorn}. In terms of the following combinations of Wilson
coefficients
\beq \label{eq:Cparallelperp}
C_{ij, \hspace{0.25mm} \parallel}^{(P, a)} = {\rm Re} \left [
  C_{ij,LL}^{(P, a)} + C_{ij,RR}^{(P, a)} \right ] \,, \qquad \quad
C_{ij, \hspace{0.25mm} \perp}^{(P, a)} = {\rm Re} \left [
  C_{ij,LR}^{(P, a)} + C_{ij,RL}^{(P, a)} \right ] \,,
\eeq
the resulting hard-scattering kernels take the form
\bea \label{eq:RSLO}
\begin{split}
  K_{q\bar{q}, \rm RS}^{(0)} &= \frac{\beta\rho}{32} \,
  \frac{\CF}{\Nc} \left [ \frac{t_1^2}{\hat s} \, C_{q \bar q,
      \hspace{0.25mm} \perp}^{(V, 8)} + \frac{u_1^2}{\hat s} \, C_{q
      \bar q, \hspace{0.25mm} \parallel}^{(V, 8)}+ m_t^2 \left ( C_{q
        \bar q, \hspace{0.25mm} \parallel}^{(V, 8)} + C_{q \bar q,
        \hspace{0.25mm} \perp}^{(V, 8)} \right )
  \right ] , \\[1mm]
  K_{t\bar{u}, \rm RS}^{(0)} &= \frac{\beta\rho}{32} \,
  \frac{\CF}{\Nc} \left [ \left ( \frac{u_1^2}{\hat s} + m_t^2 \right
    ) \left ( \frac{1}{\Nc} \hspace{0.25mm} C_{t \bar u,
        \hspace{0.25mm} \parallel}^{(V, 8)} - 2 \hspace{0.25mm} C_{t
        \bar u, \hspace{0.25mm} \parallel}^{(V, 1)} \right ) + \left (
      \frac{t_1^2}{\hat s} + m_t^2 \right ) C_{t \bar u,
      \hspace{0.25mm} \perp}^{(S, 1)} \right ]\,. \hspace{6mm}
\end{split}
\eea 
Notice that as in the SM the coefficient $K_{\bar{q} q , \rm
  RS}^{(0)}$ $\big ( K_{\bar{t} u ,\rm RS}^{(0)} \big )$ is obtained
from $K_{q \bar{q}, \rm RS}^{(0)}$ $\big ( K_{t \bar{u}, \rm RS}^{(0)}
\big )$ by simply replacing $\cos\theta$ with $-\cos\theta$.

After integrating over $\cos\theta$, one obtains the LO corrections to
the symmetric and asymmetric parts of the cross section defined in
\eq{eq:sigmatotRSLO}. In the case of the symmetric part we find
\beq \label{eq:SLONP}
\begin{split} 
  S_{u\bar{u},\rm RS}^{(0)} & = \frac{\beta\rho}{216} \, (2 + \rho) \,
  \hat s \, \left [ C_{u\bar u, \parallel}^{(V,8)} + C_{u\bar u,
      \perp}^{(V,8)} + \frac{1}{3} \hspace{0.25mm} C_{t \bar
      u, \parallel}^{(V,8)} - 2 \hspace{0.25mm} C_{t \bar
      u, \parallel}^{(V,1)} \right ] + f_S (z) \,
    \tilde C_{t \bar u}^{S}  \,, \\[1mm]
  S_{d\bar{d},\rm RS}^{(0)} & = \frac{\beta \rho}{216} \, (2 + \rho)
  \, \hat s \, \left [ C_{d\bar d, \parallel}^{(V,8)} + C_{d\bar d,
      \perp}^{(V,8)} \right ]\,,
\end{split}
\eeq 
while the asymmetric part in the partonic CM frame takes the form 
\beq \label{eq:ALONP}
\begin{split} 
  A_{u\bar{u},\rm RS}^{(0)} & = \frac{\beta^2\rho}{144} \, \hat s \,
  \left [ C_{u\bar u, \parallel}^{(V,8)} - C_{u\bar u, \perp}^{(V,8)}
    + \frac{1}{3} \hspace{0.25mm} C_{t \bar u, \parallel}^{(V,8)} - 2
    \hspace{0.25mm} C_{t \bar u, \parallel}^{(V,1)} \right ]
   + f_A (z) \, \tilde C_{t \bar
      u}^{S}  \,, \\[1mm]
  A_{d\bar{d},\rm RS}^{(0)} & = \frac{\beta^2\rho}{144} \, \hat s \,
  \left [ C_{d\bar d, \parallel}^{(V,8)} - C_{d\bar d, \perp}^{(V,8)}
  \right ]\,.
\end{split}
\eeq
Obviously, the coefficients involving down-type quarks do not receive
corrections from flavor-changing $t$-channel transitions. Notice that
in \eq{eq:SLONP} the coefficients $C_{q \bar q, \parallel}^{(V,8)}$
and $C_{q \bar q, \perp}^{(V,8)}$ enter in the combination $C_{q \bar
  q}^V \equiv \big ( C_{q \bar q, \parallel}^{(V,8)} + C_{q \bar q,
  \perp}^{(V,8)} \big )$, while in \eq{eq:ALONP} they always appear in
the form $C_{q \bar q}^A \equiv \big ( C_{q \bar q, \parallel}^{(V,8)}
- C_{q \bar q, \perp}^{(V,8)} \big )$. This feature expresses the fact
that the symmetric (asymmetric) LO cross section $\sigma_s$
($\sigma_a$) measures the product $g_V^q \hspace{0.25mm} g_V^t$ $\big
(g_A^q \hspace{0.25mm} g_A^t \big )$ of the vector (axial-vector)
parts of the couplings of the KK gluons to light quarks and top
quarks.  In order to be able to incorporate a light Higgs boson with
$m_h \ll M_{\rm KK}$ into our analysis, we have kept the full
Higgs-boson mass dependence arising from the $t$-channel
propagator.\footnote{The observant reader might wonder why we do not
  introduce form factors for the $t$-channel contribution arising from
  the $Z$-boson as well. The reason is that corrections due to
  $Z$-boson exchange turn out to be of $\ord (v^4/\Mkk^4)$. These
  effects are hence subleading and we simply ignore them in the
  following.} This dependence is described by the phase-space factors
$f_{S,A} (z)$ with $z \equiv m_h^2/m_t^2$. The analytic expressions
for $f_{S,A} (z)$ can be found in Appendix \ref{app:phasespace}. The
new Wilson coefficient $\tilde C_{t \bar u}^{S}$ is the dimensionless
counterpart of $C_{t \bar u, \perp}^{(S,1)}$.

The expressions (\ref{eq:SLONP}) and (\ref{eq:ALONP}) encode in a
model-independent way possible new-physics contributions to
$\sigma_{s,a}$ that arise from tree-level exchange of color-octet
vectors in the $s$ and $t$ channels, as well as from $t$-channel
corrections due to both new color-singlet vector and scalar
states. While this feature should make them useful in general, in the
minimal RS model based on an $SU(2)_L \times U(1)_Y$ bulk gauge
symmetry, the Wilson coefficients appearing in $S_{ij, \rm RS}^{(0)}$
and $A_{ij, \rm RS}^{(0)}$ take the following specific form. Employing
the notation of \cite{Bauer:2009cf, Casagrande:2008hr, Bauer:2008xb},
we find
\bea \label{eq:wilsonexplicit}
\begin{split}
  C^{(V, 8)}_{q \bar q, \parallel} & = -\frac{2 \pi \alpha_s}{\Mkk^2}
  \left \{ \frac{1}{L}- \sum_{a=Q,q} \left[
      (\Delta^\prime_a)_{11}+(\Delta^\prime_a)_{33}-2 L \,
      ({\widetilde \Delta}_a)_{11} \otimes({\widetilde
        \Delta}_a)_{33} \right] \right \}\,,\\
  C^{(V, 8)}_{q \bar q, \perp} & = -\frac{2 \pi \alpha_s}{\Mkk^2}
  \left \{ \frac{1}{L}- \sum_{a=Q,q} \Big [
    (\Delta^\prime_a)_{11}+(\Delta^\prime_a)_{33} \Big ] + 2L \left [
      ({\widetilde \Delta}_Q)_{11} \otimes({\widetilde \Delta}_q)_{33}
      + ({\widetilde \Delta}_q)_{11} \otimes({\widetilde
        \Delta}_Q)_{33}
    \right ] \right \}\,,\\
  C^{(V, 8)}_{t \bar u, \parallel} & = -\frac{4 \pi
    \alpha_s}{\Mkk^2}\, L\sum_{a=U,u}\Big[ ({\widetilde
    \Delta}_a)_{13}\otimes({\widetilde
    \Delta}_a)_{31}\Big]\,, \\
  C^{(V, 1)}_{t \bar u, \parallel} & = -\frac{4 \pi \alpha_e}{\Mkk^2}
  \, \frac{L}{s^2_w c^2_w} \left [ \left ( T_3^u-s^2_w Q_u \right )^2
    ({\widetilde \Delta}_U)_{13}\otimes({\widetilde \Delta}_U)_{31} +
    \left(s^2_w Q_u\right)^2 ({\widetilde \Delta}_u)_{13} \otimes
    ({\widetilde \Delta}_u)_{31} \right ] \\ & \phantom{xx} - \frac{4
    \pi \alpha_e}{\Mkk^2} \, L\, Q_u^2\sum_{a=U,u}\Big[ ({\widetilde
    \Delta}_a)_{13}\otimes({\widetilde
    \Delta}_a)_{31}\Big]\,, 
\end{split}
\eea
for $q = u, d$ and $Q = U, D$. Since the coefficient $C^{(S, 1)}_{t
  \bar u, \perp}$ is formally of ${\cal O} (v^4/\Mkk^4)$, we do not
present its explicit form. Analogous expressions with the index $1$
replaced by $2$ hold if the quarks in the initial state belong to the
second generation. Above, $\alpha_s$ ($\alpha_e$) is the strong
(electromagnetic) coupling constant, $s_w$ ($c_w$) denotes the sine
(cosine) of the weak mixing angle, whereas $T_3^u = 1/2$ and $Q_u
=2/3$ are the isospin and electric charge quantum numbers relevant for
up-type quarks. The effective couplings $(\Delta_{Q,q})_{ij}$ comprise
the overlap between KK gauge bosons and $SU(2)_L$ doublet ($Q$) or
singlet ($q$) quarks of generations $i$ and $j$.  Explicit expressions
for them can be found in \cite{Casagrande:2008hr}.  The coefficients
(\ref{eq:wilsonexplicit}) are understood to be evaluated at the KK
scale. The inclusion of RG effects, arising from the evolution down to
the top-quark mass scale, influences the obtained results only in a
minor way. Details on the latter issue can be found in
Appendix~\ref{app:RGE}. We emphasize that while the expressions for
$C^{(V, 8)}_{q \bar q, \parallel}$, $C^{(V, 8)}_{q \bar q, \perp}$,
and $C^{(V, 8)}_{t \bar u, \parallel}$ are exact, in the coefficient
$C^{(V, 1)}_{t \bar u, \parallel}$ we have only kept terms leading in
$v^2/\Mkk^2$.  The complete expression for $C^{(V, 1)}_{t \bar
  u, \parallel}$, including the subleading effects arising from the
corrections due to the mixing of fermion zero modes with their KK
excitations (that are of ${\cal O} (v^4/\Mkk^4)$), can be easily
recovered from \cite{Casagrande:2008hr}. All these corrections will
be included in our numerical analysis presented in
Section~\ref{sec:numerics}.

The expressions for the Wilson coefficients in the extended RS model
based on an $SU(2)_R \times SU(2)_L \times U(1)_X \times P_{LR}$ bulk
gauge group can be simply obtained from (\ref{eq:wilsonexplicit}) by
applying the general formalism developed in
\cite{Casagrande:2010si}. Using this formalism, one finds that the
left-handed part of the $Z$-boson contribution to $C_{t\bar
  u,\parallel}^{(V,1)}$ is enhanced by a factor of around 3, while the
right-handed contribution is protected by custodial symmetry and thus
smaller by a factor of roughly $1/L \approx 1/37$. In contrast, the
KK-gluon contributions, encoded in $C_{q \bar q,\parallel}^{(V,8)}$,
$C_{q \bar q,\perp}^{(V,8)}$, and $C_{t \bar u,\parallel}^{(V,8)}$,
remain unchanged at leading order in ${\cal O} (v^2/\Mkk^2)$. Taken
together these features imply that the predictions for the $t \bar t$
observables considered in the course of our work are rather
model-independent.

Explicit analytic expressions for the Wilson coefficients
\eq{eq:wilsonexplicit} in the ``zero-mode approximation'' (ZMA), which
at the technical level corresponds to an expansion of the exact quark
wave functions in powers of the ratio of the Higgs vacuum expectation
value (VEV) $v \approx 246 \GeV$ and the KK scale $\Mkk = \ord ({\rm
  few \ TeV})$, are given in \App{app:ZMA}. They depend on the
``zero-mode profiles'' \cite{Grossman:1999ra, Gherghetta:2000qt}
\beq
   F(c) = \mbox{sgn} \left [ \cos(\pi c) \right ]\, 
   \sqrt{\frac{1+2c}{1-\epsilon^{1+2c}}}\,,
\eeq
which are themselves functions of the bulk mass parameters
$c_{Q_i}\equiv+M_{Q_i}/k$ and $c_{q_i}\equiv-M_{q_i}/k$ that determine
the localization of the quark fields in the extra dimension. Here
$M_{Q_i}$ and $M_{q_i}$ denote the masses of the five-dimensional (5D)
$SU(2)_L$ doublet and singlet fermions, $k$ is the curvature of the 5D
anti de-Sitter (AdS$_5$) space, and $\epsilon \equiv e^{-L} \approx
10^{-16}$. 

Restricting ourselves to the corrections proportional to $\alpha_s$
and suppressing relative $\ord (1)$ factors as well as numerically
subleading terms, one finds from the results given in
(\ref{eq:wilsonuuZMA}) that the coefficient functions $S_{ij,\rm
  RS}^{(0)}$ and $A_{ij,\rm RS}^{(0)}$ introduced in \eq{eq:SLONP} and
\eq{eq:ALONP} scale in the case of the up quark like
\beq \label{eq:SALOscaling1}
\begin{split}
  S_{u\bar u, {\rm RS}}^{(0)} & \, \sim \, \frac{4 \pi
    \alpha_s}{\Mkk^2} \sum_{A = L, R }F^2 (c_{t_A}) \,, \\[1mm]
  A_{u\bar u, {\rm RS}}^{(0)} & \, \sim \, -\frac{4 \pi
    \alpha_s}{\Mkk^2} \, L \, \left \{ \prod_{q = t, u} \Big [
    F^2(c_{q_R}) - F^2(c_{q_L}) \Big ] + \frac{1}{3} \sum_{A = L, R}
    F^2(c_{t_A}) F^2(c_{u_A}) \right \} ,
\end{split}
\eeq
where $c_{t_L}\equiv c_{Q_3}$, $c_{t_R}\equiv c_{u_3}$, $c_{u_L}\equiv
c_{Q_1}$, and $c_{u_R}\equiv c_{u_1}$.

Under the natural assumptions that the bulk mass parameters of the top
and up quarks satisfy $c_{t_A} > -1/2$ and $c_{u_A} <
-1/2$,\footnote{In an anarchic approach to flavor this choice of bulk
  mass parameters is required to obtain the correct quark masses and
  mixing angles \cite{Huber:2000ie, Huber:2003tu, Agashe:2004cp,
    Bauer:2009cf}.} the relevant $F^2(c_{q_A})$ factors can be
approximated by
\beq
F^2(c_{t_A}) \approx 1 + 2 c_{t_A} \,, \qquad F^2(c_{u_A}) \approx (-1
- 2 c_{u_A}) \, e^{L \hspace{0.25mm} (2 c_{u_A} + 1)} \,, 
\eeq
with $A = L,R$. The difference of bulk mass parameters for light
quarks $(c_{u_L}-c_{u_R})$ is typically small and positive, whereas
$(c_{t_L}-c_{t_R})$ can be of $\mathcal{O}(1)$ and is usually negative
\cite{Bauer:2009cf}. Using the above approximations and expanding in
powers of $(c_{u_L} - c_{u_R})$, we find
\bea \label{eq:SALOscaling2}
\begin{split}
  S_{u\bar u, {\rm RS}}^{(0)} \, & \sim \, \frac{4 \pi
    \alpha_s}{\Mkk^2} \, 2 \left ( 1+ c_{t_L} + c_{t_R} \right ) \,, \\[1mm]
  A_{u\bar u, {\rm RS}}^{(0)} \, & \sim \, \frac{4 \pi
    \alpha_s}{\Mkk^2} \, 2 \hspace{0.25mm} L \, e^{L (1 + c_{u_L} +
    c_{u_R} )} \left ( 1 + c_{u_L} + c_{u_R} \right ) \\ & \quad \,
  \times \left \{ \left ( 2 +\frac{1}{3} \right ) L \left ( c_{t_L} -
      c_{t_R} \right ) \left ( c_{u_L} - c_{u_R} \right ) +
    \frac{1}{3} \left ( 1 + c_{t_L} + c_{t_R} \right ) \right \} ,
\end{split}
\eea 
where the symmetric function $S_{u\bar u, {\rm RS}}^{(0)}$ is entirely
due to $s$-channel KK-gluon exchange, while the contributions to the
asymmetric coefficient $A_{u\bar u, {\rm RS}}^{(0)}$ that arise from
the $s$ channel ($t$ channel) correspond to the term(s) with
coefficient $2$ ($1/3$) in the curly bracket.

The relations \eq{eq:SALOscaling2} exhibit a couple of interesting
features. We first observe that $S_{u\bar u, {\rm RS}}^{(0)}$, which
enters the RS prediction for $\sigma_s$ in \eq{eq:sigmatotRSLO}, is in
our approximation independent of the localization of the up-quark
fields and strictly positive (as long as $c_{t_A} > -1/2$). This in
turn implies an enhancement of the inclusive $t \bar t$ production
cross section which gets the more pronounced the stronger the right-
and left-handed top-quark wave functions are localized in the IR.

In contrast to $S_{u\bar u, {\rm RS}}^{(0)}$, both terms in $A_{u\bar
  u, {\rm RS}}^{(0)}$ are exponentially suppressed for UV-localized up
quarks, \ie, $c_{u_A} < -1/2$. For typical values of the bulk mass
parameters, $c_{t_L} = -0.34$, $c_{t_R} = 0.57$, $c_{u_L} = -0.63$,
and $c_{u_R} = -0.68$ \cite{Bauer:2009cf}, one finds numerically that
the first term in the curly bracket of \eq{eq:SALOscaling2}, which is
enhanced by a factor of $L$ but suppressed by the small difference
$(c_{u_L} - c_{u_R})$ of bulk mass parameters, is larger in magnitude
than the second one by almost a factor of 10. This implies that to
first order the charge asymmetry can be described by including only
the effects from $s$-channel KK-gluon exchange. Since generically $(1
+ c_{u_L} + c_{u_R}) \hspace{0.25mm} (c_{u_L} - c_{u_R}) < 0$, we
furthermore observe that a positive LO contribution to $A_{u \bar u,
  \rm RS}^{(0)}$ requires $(c_{t_L} - c_{t_R})$ to be negative, which
can be achieved by localizing the right-handed top quark sufficiently
far in the IR. To leading powers in hierarchies, one finds using the
warped-space Froggatt-Nielsen formulas given in
\cite{Casagrande:2008hr} the condition
\beq\label{eq:ctR}
c_{t_R}
\, \gtrsim \, \frac{m_t}{\sqrt{2} \hspace{0.25mm} v \left | Y_t
  \right |} - \frac{1}{2} \,,
\eeq 
where the top-quark mass is understood to be normalized at the KK
scale and $Y_t \equiv (Y_u)_{33}$ denotes the 33 element of the
dimensionless up-type quark Yukawa coupling. Numerically, this means
that for $m_t(1\TeV) = 144 \, {\rm GeV}$ and $|Y_t| =1$ values for
$c_{t_R}$ bigger than 0 lead to $A_{u\bar u ,\rm RS}^{(0)} >0$ and
thus to a positive shift in $\sigma_a$.  Taken together, it turns out
that the discussed features of $A_{u\bar u, {\rm RS}}^{(0)}$ render
the tree-level contribution to the charge asymmetry in the RS
framework tiny.\footnote{This conclusion can also be drawn from the
  statements made in \cite{Agashe:2006hk} concerning the mostly
  vector-like couplings of light quarks.}  As we will see below in our
numerical analysis, the inclusion of electroweak corrections arising
from the Born-level exchange of the $Z$ boson, of KK excitations of
both the photon and the $Z$ boson as well as of the Higgs boson, do
not change this picture qualitatively.

\subsection{Calculation of  NLO Effects}
\label{sec:NLORS}

\begin{figure}[!t]
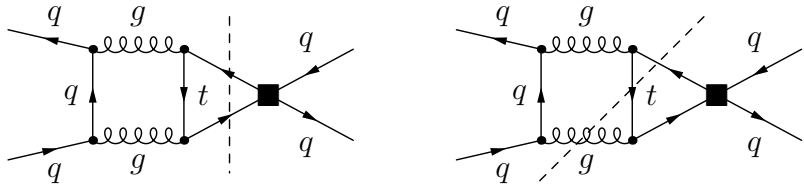

\begin{center}
\vspace{-8mm}
\mbox{
\unitlength=0.25bp
\begin{feynartspicture}(432,504)(1,1)
\FADiagram{}
\FAProp(0.,15.)(6.5,13.5)(0.,){/Straight}{-1}
\FALabel(3.59853,15.2803)[b]{$q$}
\FAProp(0.,5.)(6.5,6.5)(0.,){/Straight}{1}
\FALabel(3.59853,4.71969)[t]{$q$}
\FAProp(20.,10.)(13.5,13.5)(0.,){/Straight}{1}
\FAProp(20.,10.)(13.5,6.5)(0.,){/Straight}{-1}
\FAProp(6.5,13.5)(6.5,6.5)(0.,){/Straight}{-1}
\FALabel(5.43,10.)[r]{$q$}
\FAProp(6.5,13.5)(13.5,13.5)(0.,){/Cycles}{0}
\FALabel(10.,15.27)[b]{$g$}
\FAProp(6.5,6.5)(13.5,6.5)(0.,){/Cycles}{0}
\FALabel(10.,5.43)[t]{$g$}
\FAProp(13.5,13.5)(13.5,6.5)(0.,){/Straight}{1}
\FALabel(14.57,10.)[l]{$t$}
\FAVert(6.5,13.5){0}
\FAVert(6.5,6.5){0}
\FAVert(13.5,13.5){0}
\FAVert(13.5,6.5){0}
\FALabel(20.0,10.0)[c]{$\blacksquare$}
\FAProp(20.0,10.0)(26.5,13.5)(0.,){/Straight}{-1}
\FAProp(20.0,10.0)(26.5,6.5)(0.,){/Straight}{1}
\FALabel(22.9015,13.2803)[b]{$q$}
\FALabel(22.9015,6.71969)[t]{$q$}
\FAProp(17.,16.)(17.,4.0)(0.,){/ScalarDash}{0}
\end{feynartspicture}
}
\hspace{16mm}
\mbox{
\unitlength=0.25bp
\begin{feynartspicture}(432,504)(1,1)
\FADiagram{}
\FAProp(0.,15.)(6.5,13.5)(0.,){/Straight}{-1}
\FALabel(3.59853,15.2803)[b]{$q$}
\FAProp(0.,5.)(6.5,6.5)(0.,){/Straight}{1}
\FALabel(3.59853,4.71969)[t]{$q$}
\FAProp(20.,10.)(13.5,13.5)(0.,){/Straight}{1}
\FAProp(20.,10.)(13.5,6.5)(0.,){/Straight}{-1}
\FAProp(6.5,13.5)(6.5,6.5)(0.,){/Straight}{-1}
\FALabel(5.43,10.)[r]{$q$}
\FAProp(6.5,13.5)(13.5,13.5)(0.,){/Cycles}{0}
\FALabel(10.,15.27)[b]{$g$}
\FAProp(6.5,6.5)(13.5,6.5)(0.,){/Cycles}{0}
\FALabel(10.,5.43)[t]{$g$}
\FAProp(13.5,13.5)(13.5,6.5)(0.,){/Straight}{1}
\FALabel(14.57,10.)[l]{$t$}
\FAVert(6.5,13.5){0}
\FAVert(6.5,6.5){0}
\FAVert(13.5,13.5){0}
\FAVert(13.5,6.5){0}
\FALabel(20.0,10.0)[c]{$\blacksquare$}
\FAProp(20.0,10.0)(26.5,13.5)(0.,){/Straight}{-1}
\FAProp(20.0,10.0)(26.5,6.5)(0.,){/Straight}{1}
\FALabel(22.9015,13.2803)[b]{$q$}
\FALabel(22.9015,6.71969)[t]{$q$}
\FAProp(19.,16.)(6.,3.)(0.,){/ScalarDash}{0}
\end{feynartspicture}
}
\end{center}
\vspace{-1.5cm}
\begin{center}
  \parbox{15.5cm}{\caption{\label{fig:RSLoop} Representative diagrams
      contributing to the forward-backward asymmetry in $t \bar t$
      production at NLO. The two-particle (three-particle) cut
      (represented by a dashed line) corresponds to the interference
      of $q \bar q \to t \bar t$ ($q \bar q \to t \bar t
      \hspace{0.25mm} g$) with $Q_{q \bar q, AB}^{(V, 8)}$. The
      insertion of the effective operator is indicated by a black
      square. The SM contribution is simply obtained by replacing the
      operator by $s$-channel gluon exchange.}}
\end{center}
\end{figure}

In models with small axial-vector couplings to light quarks and no
significant FCNC effects in the $t$ channel, the charge-asymmetric
cross section $\sigma_a$ is suppressed at LO. As we will show in the
following, this suppression can be evaded by going to NLO, after
paying the price of an additional factor of $\alpha_s/(4 \pi)$. In
order to understand how the LO suppression is lifted at the loop
level, it is useful to recall the way in which the charge asymmetry
arises in the SM. Since QCD is a pure vector theory, the lowest-order
processes $q \bar q \to t \bar t$ and $gg \to t \bar t$, which are of
$\ord (\alpha_s^2)$ do not contribute to $\AFBt$. However, starting at
$\ord (\alpha_s^3)$, quark-antiquark annihilation $q\bar{q} \to t \bar
t \hspace{0.5mm} (g)$, as well as flavor excitation $qg \to q t \bar
t$ receive charge-asymmetric contributions \cite{Kuhn:1998jr,
  Kuhn:1998kw}, while gluon fusion $gg \to t \bar t \hspace{0.5mm}
(g)$, remains symmetric to all orders in perturbation theory. Charge
conjugation invariance can be invoked to show that, as far as the
virtual corrections to $q\bar{q}\rightarrow t\bar{t}$ are concerned,
only the interference between the lowest-order and the QCD box graphs
contributes to the asymmetry at NLO. Similarly, for the
bremsstrahlungs (or real) contributions, only the interference between
the amplitudes that are odd under the exchange of $t$ and $\bar t$
furnishes a correction. Since the axial-vector current is even under
this exchange, the NLO contribution to the asymmetry arises solely
from vector-current contributions. These features imply that at NLO
the charge-asymmetric cross section is proportional to the $d_{abc}^2
= \left (2 \hspace{0.25mm} {\rm Tr} \left ( \{ T^a, T^b \} T^c \right
  ) \right )^2$ terms \cite{Kuhn:1998jr, Kuhn:1998kw} that originate
from the interference of both the one-loop box and $t \bar t g$ final
states with the tree-level quark-antiquark annihilation diagram. The
relevant Feynman graphs are obtained from the ones shown in
\Fig{fig:RSLoop} by replacing the operator insertion by an $s$-channel
gluon exchange. The QCD expression for $\sigma_a$ can be derived from
the analogous quantity in the electromagnetic process $e^+ e^- \to
\mu^+ \mu^-$ \cite{Berends:1973fd, Berends:1982dy} by a suitable
replacement of the QED coupling and the electromagnetic
charges. Explicit formulas for the asymmetric contributions to the
$t\bar{t}$ production cross section in QCD are given in
\cite{Kuhn:1998kw}. Contributions from flavor excitation are
negligibly small at the Tevatron and will not be taken into account in
the following.

The main lesson learned from the way the charge asymmetry arises in
QCD is that beyond LO vector couplings alone are sufficient to
generate non-vanishing values of $\AFBt$. In the case of the EFT
(\ref{eq:Leff}) this means that cut diagrams like the ones shown in
\Fig{fig:RSLoop}, can give a sizable contribution to the charge
asymmetry if the combination $C_{q \bar q}^V = \big ( C_{q \bar
  q, \parallel}^{(V,8)} + C_{q \bar q, \perp}^{(V,8)} \big )$ of
Wilson coefficients is large enough.  In fact, from \eq{eq:SLONP},
\eq{eq:ALONP}, and \eq{eq:SALOscaling2} it is not difficult to
convince oneself that in the case of the RS model NLO corrections to
$\sigma_a$ should dominate over the LO ones, if the
condition\footnote{This inequality should be considered only as a
  crude approximation valid up to a factor of $\ord (1)$.}
\beq \label{eq:naive} 
\frac{\alpha_s}{4\pi} \, (1+c_{t_L}+c_{t_R}) \, \gtrsim \, L \, e^{L
  (1+ c_{u_L}+c_{u_R}) }
\eeq 
is fulfilled. For example, employing $c_{t_L} = -0.34$, $c_{u_L} =
-0.63$, and $c_{u_R} = -0.68$, the above formula tells us that for
$c_{t_R} = 0.57$ the NLO contributions are bigger than the LO
corrections by a factor of roughly 25. This suggests that it might be
possible to generate values of $\AFBt$ that can reach the percent
level with typical and completely natural choices of
parameters. Notice that in contrast to QCD, in the RS framework the
Feynman graphs displayed in \Fig{fig:RSLoop} are not the only sources
of charge-asymmetric contributions. Self-energy, vertex, and
counterterm diagrams will also lead to an asymmetry.\footnote{Box
  diagrams involving the virtual exchange of one zero-mode and one KK
  gluon potentially also give a contribution to $\AFBt$ at NLO. We do
  not include such effects here.} However, these corrections are, like
the Born-level contribution, all exponentially suppressed by the UV
localization of the light-quark fields (and the small axial-vector
coupling of the light quarks for what concerns the contributions from
the operators $Q_{q\bar q, AB}^{(V,8)}$). Compared to the tree-level
corrections, these contributions are thus suppressed by an additional
factor of $\alpha_s/(4 \pi)$, so that they can be ignored for all
practical purposes.

\begin{figure}[!t]
\begin{center}
\includegraphics[width=7.5cm]{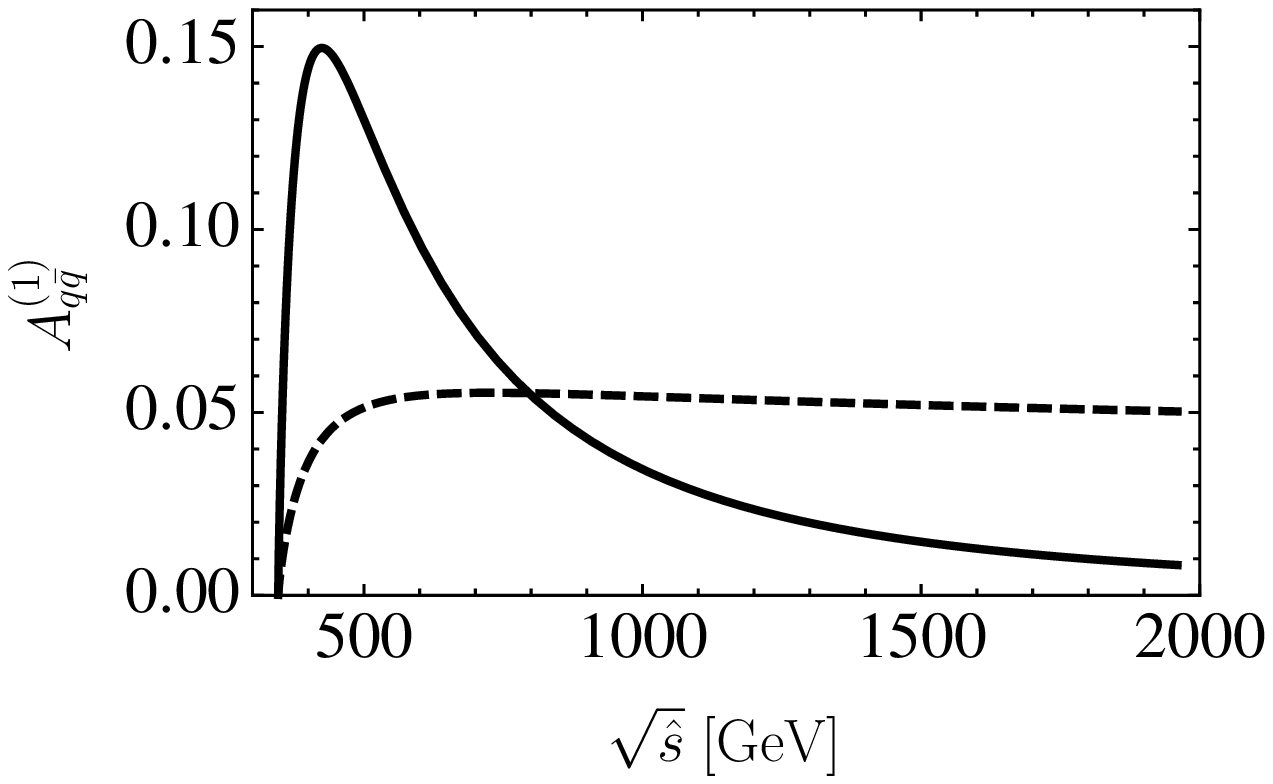}
\quad 
\includegraphics[width=7.25cm]{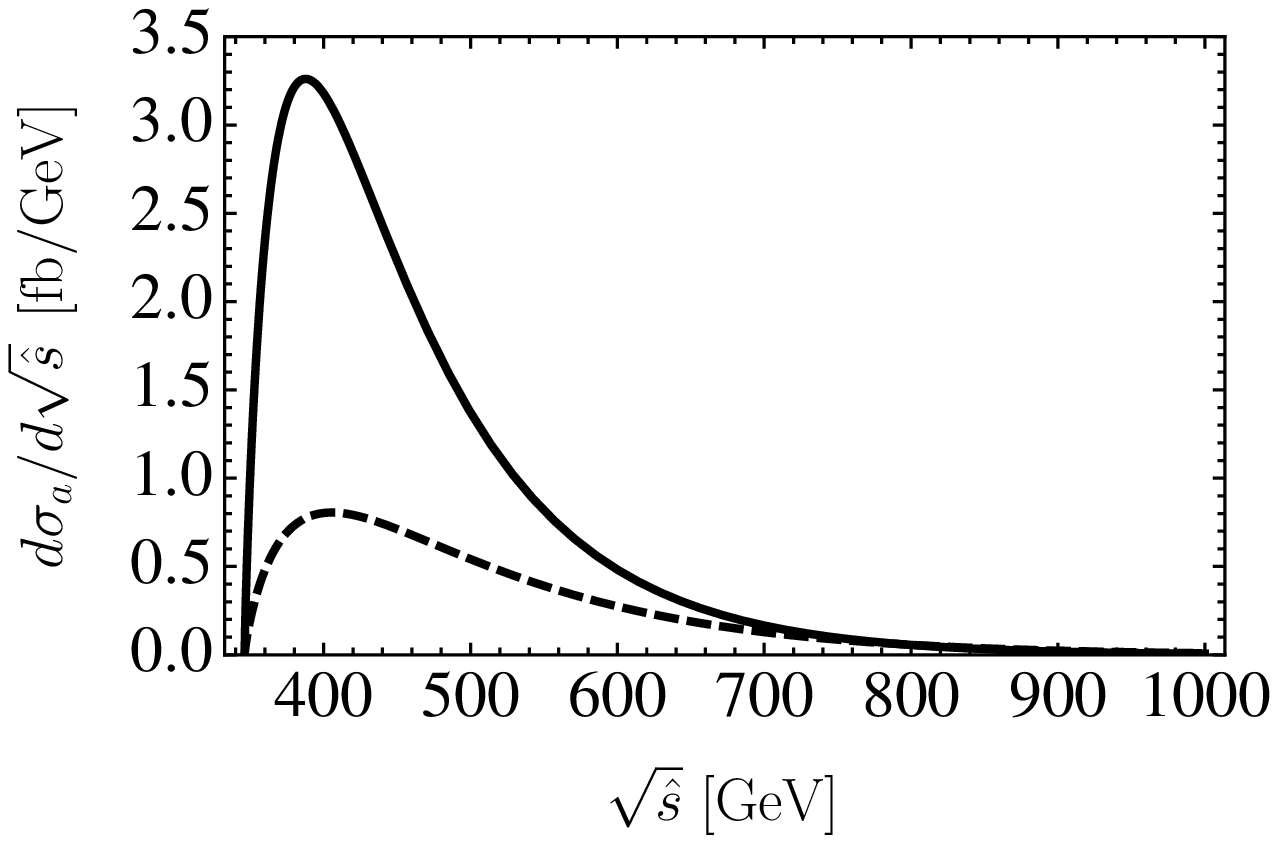}
\end{center}
\vspace{-7.5mm}
\begin{center}
  \parbox{15.5cm}{\caption{\label{fig:A1qq} The asymmetric coefficient
      $A^{(1)}_{q\bar q}$ (left panel) and the differential hadronic
      asymmetry $d\sigma_a/d\sqrt{\hat s}$ (right panel) as functions
      of $\sqrt{\hat s}$ in the SM (solid lines) and the RS model
      (dashed lines). For presentational purposes the shown RS
      contributions have been obtained using the fictitious value
      $C_{q\bar q}^V=10/\TeV^{2}$. See text for further details.}}
\end{center}
\end{figure}

The above explanations should be motivation enough to perform a
calculation of $\AFBt$ in the RS model beyond LO including
the graphs depicted in \Fig{fig:RSLoop}. After integrating over $\cos
\theta$, we obtain in the partonic CM frame ($q \bar q = u \bar u,
d\bar d$) 
\beq\label{eq:NPNLO}
A^{(1)}_{q\bar{q},\rm RS} = \frac{\hat{s}}{16\pi\alpha_s}
\hspace{0.5mm} C_{q\bar q}^V \hspace{0.5mm} A^{(1)}_{q\bar{q}}\,,
\eeq 
where $A^{(1)}_{q\bar{q}}$ denotes the NLO asymmetric SM coefficient,
normalized according to (\ref{eq:Cijexp}).  This function can be
described through a parametrization which is accurate to the permille
level. The result of our fit reads
\beq 
  A^{(1)}_{q\bar{q}} = \frac{\alpha_s\,d_{abc}^2}{16N_c^2} \;
  5.994\,\beta\rho\,\Big [1 + 17.948\,\beta - 20.391\,\beta^2 +
  6.291\,\beta^3 + 0.253\,\ln\left (1-\beta \right )\Big ]\,, 
\eeq 
where $N_c = 3$ and $d_{abc}^2 = \left (\Nc^2-1 \right ) \left
  (\Nc^2-4 \right )/\Nc = 40/3$. It has been obtained by integrating
the expressions for the charge-asymmetric contributions to the
differential $t \bar t$ production cross section given in
\cite{Kuhn:1998kw} over the relevant phase space.\footnote{The
  numerical integration has been performed using the Vegas Monte Carlo
  algorithm implemented in the {\tt CUBA} library \cite{Hahn:2004fe}}
In the left panel of \Fig{fig:A1qq} we show $A^{(1)}_{q\bar{q}}$ as a
function of $\sqrt{\hat s}$, employing $m_t = 173.1 \, {\rm GeV}$
\cite{top:2009ec} and $\alpha_s (m_t) = 0.126$. We clearly see that
the SM distribution $A^{(1)}_{q\bar{q}}$ (solid curve) peaks at around
$\sqrt{\hat s} \approx 420 \, {\rm GeV}$, \ie, relatively close to the
$t \bar t$ threshold. The NLO RS contribution $A^{(1)}_{q\bar{q},\rm
  RS}$ (dashed curve) does not exhibit such a drop-off at large
$\sqrt{\hat s}$ due to the additional factor of $\hat s$ in
(\ref{eq:NPNLO}). Since the quark luminosities $\ff_{ij}(\hat
s/s,\mu_f)$ fall off strongly with $\hat s$, behaving roughly like
$1/\hat s^2$, the integrated asymmetry in (\ref{eq:sigmatotRSLO}) is
saturated well before the upper integration limit $s$ is reached. This
can be seen from the right panel in \Fig{fig:A1qq}, where we multiplied 
the coefficients $A^{(1)}_{q\bar q}$ and $A^{(1)}_{q\bar{q},\rm RS}$
with the up-quark PDFs $\ff_{u\bar u}(\hat s/s,\mu_f)$.

\section{Numerical Analysis}
\label{sec:numerics}

The Wilson coefficients appearing in the effective Lagrangian
(\ref{eq:Leff}) are constrained by the measurements of the
forward-backward asymmetry $\AFBt$, the total cross section $\sigtot$,
and the $t\bar t$ invariant mass spectrum $\dsig$. The latest result
for $\AFBt$ has already been given in (\ref{eq:AFBexp}) and the most
recent Tevatron results ($\sqrt{s} = 1.96 \, {\rm TeV}$) for the
remaining measurements of interest read \cite{CDFnotetot,
  Bridgeman:2008zz, Aaltonen:2009iz}
\begin{gather} 
  (\sigtot)_{\rm exp} = (7.50 \pm 0.31_{\rm stat.} \pm 0.34_{\rm
    syst.} \pm 0.15_{\rm lumi.})\,\rm pb\,, \nonumber
  \\ \label{eq:EXPss} \\[-4mm]
  \left (\frac{d \sigtot}{d \Mtt} \right )_{\rm exp}^{\Mtt \, \in \,
    [800, 1400] \GeV} = (0.068 \pm 0.032_{\rm stat.} \pm 0.015_{\rm
    syst.}  \pm 0.004_{\rm lumi.})\,\frac{\rm fb}{\GeV}\,. \nonumber
\end{gather}
Here the quoted individual errors are of statistical and systematic
origin, and due to the luminosity uncertainty, respectively. Notice
that in the case of the $t\bar t$ invariant mass spectrum, we have
restricted our attention to the last bin of the available CDF
measurement, \ie, $\Mtt \in \, [800, 1400] \GeV$, which is most
sensitive to the presence of new degrees of freedom with masses in the
TeV range.

The above results should be compared to the predictions obtained in
the SM supplemented by the dimension-six operators
(\ref{eq:Leff}). Ignoring tiny contributions related to the
(anti)strange-, (anti)charm-, and (anti)bottom-quark content of the
proton (antiproton), we find in terms of the dimensionless
coefficients $\tilde C_{q \bar q}^V \equiv 1\, {\rm TeV}^2 \, C_{q
  \bar q}^V$ and $\tilde C_{t \bar u}^V \equiv 1\, {\rm TeV}^2 \, \big
( 1/3 \hspace{0.5mm} C_{t \bar u, \parallel}^{(V,8)} - 2
\hspace{0.25mm} C_{t \bar u, \parallel}^{(V,1)} \big )$ the following
expressions
\begin{gather} 
  (\sigtot)_{\rm RS} = \left [ 1 + 0.053 \hspace{0.5mm} \big ( \tilde C_{u \bar
      u}^V + \tilde C_{t \bar u}^{V} \big ) - 0.612\, \tilde C_{t \bar
      u}^{S} + 0.008 \, \tilde C_{d \bar d}^V \hspace{0.25mm} \right ]
  \left ( 6.73^{+0.52}_{-0.80} \right ) {\rm pb} \,, \nonumber \\ 
  \label{eq:EFTss} \\[-4mm]
  \left (\frac{d \sigtot}{d \Mtt} \right )^{\Mtt \, \in \, [800, 1400]
    \GeV}_{ \rm RS} = \left [ 1 + 0.33 \hspace{0.5mm} \big ( \tilde C_{u \bar
      u}^V + \tilde C_{t \bar u}^{V} \big ) - 0.81 \, \tilde C_{t \bar
      u}^{S} + 0.02 \, \tilde C_{d \bar d}^V \hspace{0.25mm} \right ]
  \left ( 0.061^{+0.012}_{-0.006} \right ) \frac{\rm fb}{\rm GeV} \,,
  \hspace{4mm} \nonumber
\end{gather}
where all Wilson coefficients are understood to be evaluated at
$m_t$. The numerical factors multiplying $\tilde C_{t \bar u}^{S}$
correspond to a Higgs mass of $m_h = 115 \, {\rm GeV}$, which we will
adopt in the following.  The RG evolution of the Wilson coefficients
from $\Mkk$ to $m_t$ is achieved with the formulas given in
Appendix~\ref{app:RGE}. The dependence of $\sigtot$ and $\dsig$ on
$\tilde C_{ij}^P$ has been obtained by convoluting the kernels
\eq{eq:SLONP} with the parton luminosities $\ff_{ij}(\hat s/s,\mu_f)$
by means of the charge-symmetric analog of formula
\eq{eq:sigmatotRSLO}, using {\tt MSTW2008LO}
PDFs~\cite{Martin:2009iq} with renormalization and factorization
scales fixed to the reference point $\mu_r = \mu_f=m_t=173.1\,
\GeV$. The corresponding value of the strong coupling constant is
$\alpha_s(M_Z)=0.139$, which translates into $\alpha_s(m_t)=0.126$
using one-loop RG running. The total cross section and $t \bar t$
invariant mass distribution in the SM have been calculated at NLO
\cite{Nason:1987xz} with the help of {\tt MCFM} \cite{MCFM},
employing {\tt MSTW2008NLO} PDFs along with $\alpha_s(M_Z) = 0.120$, 
corresponding to $\alpha_s (m_t) = 0.109$ at two-loop accuracy. The
given total SM errors represent the uncertainty due to the variation
$\mu_r = \mu_f \in [m_t/2, 2 \hspace{0.25mm} m_t]$ as well as PDF
errors within their 90\% confidence level~(CL) limits, after combining
the two sources of error in quadrature. Notice that within errors our
SM prediction for $\sigtot$ is in good agreement with recent
theoretical calculations, that include effects of logarithmically
enhanced NNLO terms \cite{Ahrens:2010zv, Cacciari:2008zb,
  Kidonakis:2008mu, Langenfeld:2009wd, Ahrens:2009uz}.

The forward-backward asymmetry $\AFBt$ as given in (\ref{eq:AFBexp})
is measured in the $p\bar p$ frame, while (\ref{eq:chargeas}),
(\ref{eq:fbas}), and (\ref{eq:sigmatotRSLO}) apply in the partonic CM
frame. The transformation from the partonic CM into the $p\bar p$
frame corresponds to a mere change of integration boundaries of the
scattering angle $\cos\theta$. In order to calculate the asymmetric
contribution to the cross section in the $p\bar p$ frame,
$\sigma_a^{p\bar p}$,  we employ at Born level,
\beq \label{eq:sigmaalab}
\sigma_a^{p\bar p} =
\frac{\alpha_s}{m_t^2}\sum_{i,j}\int_{4m_t^2/s}^{1} d\tau
\int_{\tau}^1 \frac{dx}{x}\,f_{i/p}(x,\mu_f)\,f_{j/\bar
  p}(\tau/x,\mu_f) \, A_{ij}^{p\bar p}(x,\tau,\mu_f)\,,
\eeq
where $\tau\equiv \hat s/s$ and 
\beq \label{eq:Aijppbar}
A_{ij}^{p\bar p}(x,\tau,\mu_f) \equiv \int_{c(x,\tau)}^1 
d\cos\theta\, K_{ij}(\rho,\cos\theta,\mu_f) - \int_{-1}^{c(x,\tau)} 
 d\cos\theta\, K_{ij}(\rho,\cos\theta,\mu_f)\,,
\eeq
with 
\beq \label{eq:cxtau}
 c(x,\tau) \equiv 
  \frac{1}{\beta}\frac{x^2-\tau}{x^2 + \tau}\,,
\eeq
and the hard-scattering kernels $K_{ij}(\rho,\cos\theta,\mu_f)$ have
been introduced in (\ref{eq:Cijexp}). Beyond LO the phase-space
integration is more involved. For convenience we thus give the
reduction factors $R\equiv \sigma_a^{p\bar p}/\sigma_a$ that are
needed to convert the SM as well as the EFT results of the
forward-backward asymmetry from the partonic CM to the $p \bar p$
frame.  In the SM we find $R_{\rm SM} = 0.64$, while the ratios
necessary to calculate the contributions arising from the various
effective operators are given by $R_{u \bar u}^V = 0.73$, $R_{d \bar
  d}^V = 0.72$, $R_{t \bar u}^S = -1.78$, $R_{u \bar u}^A = 0.58$, and
$R_{d \bar d}^A = 0.56$.  These numbers correspond to {\tt MSTW2008LO}
PDFs with the renormalization and factorization scales set to the
reference point mentioned above.
 
With all this at hand, we are now in a position to give the
forward-backward asymmetry in the laboratory frame. Normalizing the
result for $\sigma_a^{p \bar p}$ to $\sigma_s$ calculated at
NLO,\footnote{Using {\tt MSTW2008} PDFs and $\mu_r = \mu_f = m_t =
  173.1 \, {\rm GeV}$, we obtain in the SM the symmetric cross
  sections $(\sigma_s)_{\rm LO} = 6.66 \, {\rm pb}$ and
  $(\sigma_s)_{\rm NLO} = 6.73 \, {\rm pb}$ from {\tt MCFM}. Since
  these results differ by only 1\%, the central value of $\AFBt$ does
  essentially not depend on whether the LO or the NLO cross section is
  used to normalize (\ref{eq:AFBEFT}).} we find the following
expression
\beq \label{eq:AFBEFT} 
(\AFBt)^{p \bar p}_{\rm RS} = \left [ \frac{1 + 0.22
    \hspace{0.5mm} \big ( \tilde C_{u \bar u}^A + \tilde C_{t \bar
      u}^V \big ) +0.72 \hspace{0.25mm} \tilde C_{t \bar u}^S + 0.03
    \hspace{0.25mm} \tilde C_{d \bar d}^A + 0.034 \hspace{0.25mm}
    \tilde C_{u \bar u}^V + 0.005 \hspace{0.25mm} \tilde C_{d \bar
      d}^V}{1 + 0.053 \hspace{0.5mm} \big ( \tilde C_{u \bar u}^V +
    \tilde C_{t \bar u}^V \big ) -0.612 \hspace{0.25mm} \tilde C_{t
      \bar u}^S + 0.008 \hspace{0.25mm} \tilde C_{d \bar d}^V } \right
] \! \left ( 5.6^{+0.8}_{-1.0} \right ) \% \,,
\eeq  
where all coefficient functions should be evaluated at the scale
$m_t$. The central value of our SM prediction has been obtained by
integrating the formulas given in \cite{Kuhn:1998kw} over the relevant
phase space (see (\ref{eq:sigmaalab}) to (\ref{eq:cxtau})), weighted
with {\tt MSTW2008LO} PDFs with the unphysical scales fixed to
$m_t$. It is in good agreement with (\ref{eq:AFBSM}) as well as the
findings of \cite{Almeida:2008ug, Ahrens:2010zv}. Unlike
\cite{Antunano:2007da}, we have chosen not to include electroweak
corrections to the forward-backward asymmetry in the central value of
(\ref{eq:AFBEFT}). Such effects have been found in
\cite{Antunano:2007da, Bernreuther:2010ny} to enhance the $t \bar t$
forward-backward asymmetry by around $9\%$ to $4\%$ depending on
whether only mixed electroweak-QCD contributions or also purely
electroweak corrections are included. To account for the uncertainty
of our SM prediction due to electroweak effects we have added in
quadrature an error of $5\%$ to the combined scale and PDF
uncertainties.

\begin{table}
\begin{center}
\begin{tabular}{|c|c|c|c|c|c|c|c|}
  \hline
  $c_{t_L}$ & $c_{t_R}$ & $\tilde C_{u\bar u}^V/\alpha_s$ & 
  $\tilde C_{u\bar u}^A/\alpha_s$ & 
  $\tilde C_{d\bar d}^V/\alpha_s$ & 
  $\tilde C_{d\bar d}^A/\alpha_s$ & 
  $\tilde C_{t\bar u}^V/\alpha_s$ & 
  $\tilde C_{t \bar u}^S$ \\
  \hline
  $-0.41$ & $0.09$ & $4.50$ & $0.71 \cdot 10^{-2}$ & $0.68$ & 
   $-1.40 \cdot 10^{-3}$ & $-1.35 \cdot 10^{-4}$ & $8.2 \cdot 10^{-7}$ \\
  $-0.47$ & $0.48$ & $4.95$ & $0.22 \cdot 10^{-2}$ & $0.27$ & 
  $-0.03 \cdot 10^{-3}$ & $-0.70 \cdot 10^{-4}$ & $4.1 \cdot 10^{-7}$ \\
  $-0.49$ & $0.90$ & $5.31$ & $1.79 \cdot 10^{-2}$ & $0.08$ & 
  $-0.64 \cdot 10^{-3}$ & $-2.45 \cdot 10^{-4}$ & $122 \cdot 10^{-7}$ \\
  \hline
\end{tabular}
\end{center}
\caption{\label{tab:WC} 
  Results for the Wilson coefficients corresponding to three different 
  parameter points. The numbers shown correspond to the RS model with 
  $SU(2)_L \times U(1)_Y$ bulk gauge symmetry and brane-localized Higgs 
  sector. The coefficients in the first five columns scale as 
  $(1 \, {\rm TeV}/\Mkk)^2$ while the one in the last column behaves as 
  $(1 \, {\rm TeV}/\Mkk)^4$. Further details are given in the text. 
}
\end{table}

In order to investigate the importance of the different contributions
entering the RS predictions (\ref{eq:EFTss}) and (\ref{eq:AFBEFT}) for
the $t \bar t$ observables, we have calculated the relevant Wilson
coefficients at the KK scale for typical sets of bulk mass parameters
and anarchic Yukawa couplings $Y_{u,d}$ (\ie, non-hierarchical
matrices with ${\cal O}(1)$ complex elements). In \Tab{tab:WC} we
present our numerical results for the coefficient functions for three
assorted parameter points that reproduce the observed quark masses as
well as the angles and the CP-violating phase in the quark mixing
matrix within errors. To keep the presentation simple, we show in the
table only the values of the left- and right-handed top-quark bulk
mass parameters $c_{t_L}$ and $c_{t_R}$. The numerical values for the
remaining bulk mass parameters and Yukawa matrices, specifying the
three parameter points completely, are relegated to
\App{app:points}. We emphasize that the magnitudes of the shown
results are generic predictions in the allowed parameter space and do
not reflect a specific choice of model parameters. From the numbers
given in the table, we see that the ratios of magnitudes of the Wilson
coefficients are given by $|\tilde C_{q \bar q}^A|/|\tilde C_{q \bar
  q}^V| = {\cal O} (10^{-3})$, $|\tilde C_{t \bar u}^V|/|\tilde C_{u
  \bar u}^V| = {\cal O} (10^{-5})$, and $|\tilde C_{t \bar
  u}^S|/|\tilde C_{u \bar u}^V| = {\cal O} (10^{-6})$. For what
concerns the size of the corrections due to flavor-changing currents
in the $t$ channel (encoded in $\tilde C_{t \bar u}^V$ and $\tilde
C_{t \bar u}^S$), we mention that in the RS model based on an $SU(2)_L
\times U(1)_Y$ bulk gauge group, the ratio of neutral electroweak
gauge boson (Higgs-boson) to KK-gluon effects is roughly $1/3$ (on
average $1/50$). In the RS variant with extended $SU(2)_R$ symmetry
and custodial protection of the $Z b_L \bar b_L$ vertex, one finds a
very similar pattern. The quoted numbers imply that the predictions
for the $t \bar t$ observables considered in our work are fairly
model-independent, as they do not depend sensitively on the exact
realizations of neither the electroweak gauge, nor the fermionic, nor
the Higgs sector.

\begin{figure}[!t]
\begin{center}
\includegraphics[width=7.5cm]{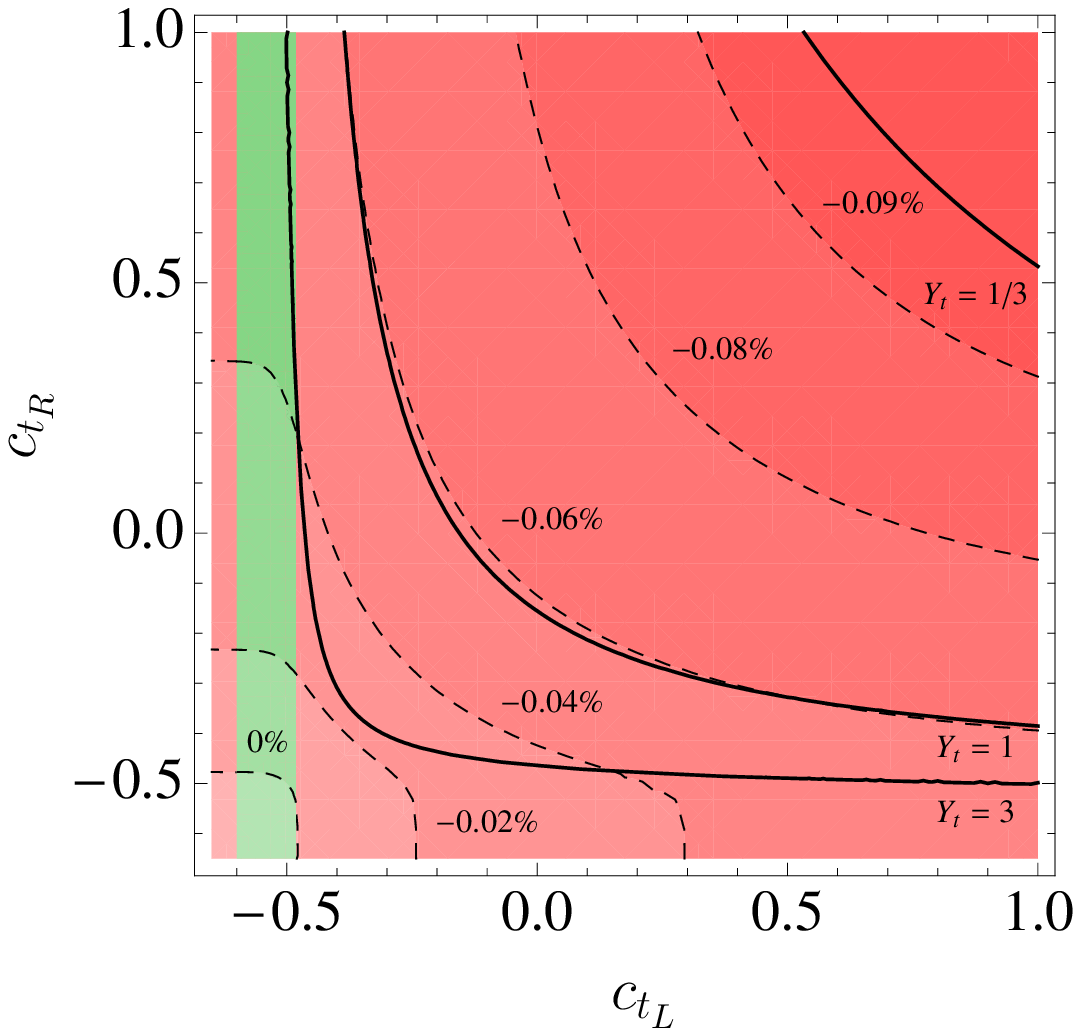}
\qquad 
\includegraphics[width=7.5cm]{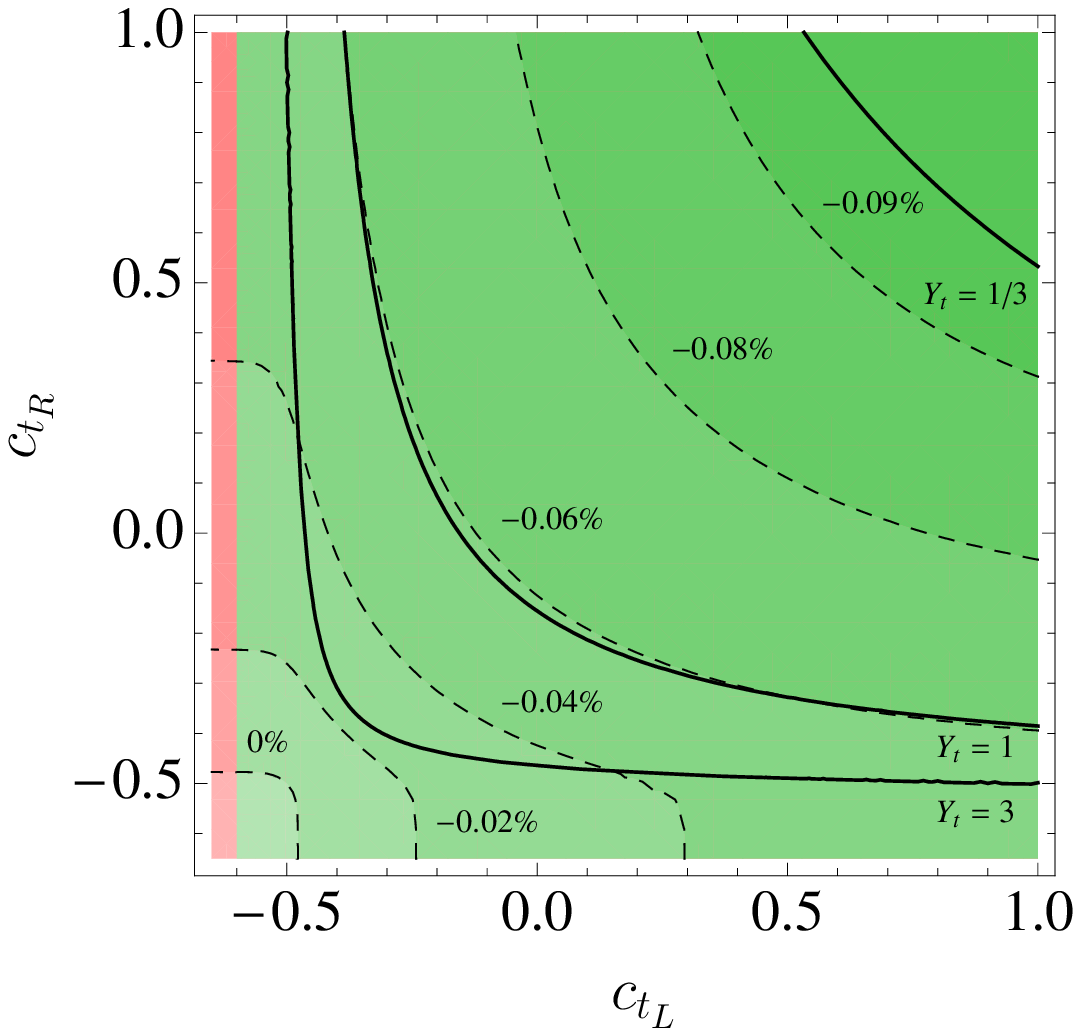}
\end{center}
\vspace{-7.5mm}
\begin{center}
  \parbox{15.5cm}{\caption{\label{fig:RSpredictions} Size of the
      absolute correction to $(\AFBt)_{\rm RS}^{p \bar p}$ in the
      $c_{t_L}$--$c_{t_R}$ plane for a KK scale of $1 \, {\rm
        TeV}$. The solid lines indicate the value of $Y_t$ necessary
      to reproduce the measured mass of the top quark.  In the left
      (right) panel only the parameter region of the minimal
      (extended) RS model displayed in green satisfies the constraints
      imposed by the $Z \to b \bar b$ ``pseudo observables''. See text
      for further details.}}
\end{center}
\end{figure}

Focusing on the numerical dominant corrections arising from
$s$-channel KK-gluon exchange, we see from \Tab{tab:WC} that $\tilde
C_{u \bar u}^V$ and $\tilde C_{u \bar u}^A$ are a factor of a few
larger in magnitude than their counterparts involving down
quarks. Since the latter coefficients are suppressed in the total
cross section (last bin of the $t \bar t$ invariant mass spectrum)
relative to the coefficients involving up quarks by the small ratio of
quark luminosities $\ff_{d\bar d} \left (0.04 \right )/\ff_{u\bar u}
\left (0.04 \right ) \approx 1/5$ ($\ff_{d\bar d} \left (0.17 \right
)/\ff_{u\bar u} \left (0.17 \right ) \approx 1/15$), the numerical
impact of $\tilde C_{d \bar d}^V$ in (\ref{eq:EFTss}) is
negligible. In practice, we find that the relevant ratio $(0.008
\hspace{0.5mm} \tilde C_{d \bar d}^V)/(0.053 \hspace{0.5mm} \tilde
C_{u \bar u}^V)$ $\big ( (0.02 \hspace{0.5mm} \tilde C_{d \bar
  d}^V)/(0.33 \hspace{0.5mm} \tilde C_{u \bar u}^V) \big )$ amounts to
less than $2.3\%$ ($1.0\%$) for the considered parameter points. In
the following, we will therefore restrict our attention to the
coefficients $\tilde C_{u \bar u}^{V, A}$ that render by far the
largest contributions to the $t \bar t$ observables in the RS
model. From \Tab{tab:WC} we first observe that $\tilde C_{u \bar
  u}^{V}$ grows with increasing $(c_{t_L} + c_{t_R})$, \ie, when the
top quark is localized more strongly in the IR (as expected from
\eq{eq:SALOscaling2} and \eq{eq:ctR}). A similar trend in terms of
$c_{t_R}$, though less pronounced, is also visible in the case of
$\tilde C_{u \bar u}^{A}$. The numbers given in the table furthermore
confirm our qualitative findings from \Sec{sec:LORS} of strongly
suppressed axial-vector couplings, $|\tilde C_{u\bar u}^A|/|\tilde
C_{u\bar u}^V| \ll 1$. Inserting the numerical values of $\tilde C_{u
  \bar u}^{V}$ and $\tilde C_{u \bar u}^{A}$ into the numerator of
(\ref{eq:AFBEFT}), we see that also our third expectation (made in
\Sec{sec:NLORS}) that in the RS model the NLO contributions to $\AFBt$
arising from $\tilde C_{u \bar u}^{V}$ are bigger than the LO
corrections stemming from $\tilde C_{u \bar u}^{A}$, in fact, holds
true. Numerically, we find that the vector-current contributions to
the asymmetry are typically larger by about a factor of 100 than the
corrections due to the axial-vector current.

While this strong enhancement looks promising at first sight, a closer
inspection of (\ref{eq:AFBEFT}) shows that in the ratio of the
asymmetric and symmetric cross sections the effects of $\tilde C_{u
  \bar u}^{V}$ tend to cancel. Since both $\sigma_a^{p \bar p}$ and
$\sigma_s$ are enhanced for $\tilde C_{u \bar u}^V > 0$, but the
dependence of $\sigma_s$ on $\tilde C_{u \bar u}^V$ is stronger than
the one of $\sigma_a^{p \bar p}$, positive values of $\tilde C_{u \bar
  u}^V$ will effectively lead to a reduction and not to an enhancement
of the $t \bar t$ forward-backward asymmetry as envisioned in
\Sec{sec:NLORS}. Given that $\tilde C_{u \bar u}^V > 0$ is a robust
prediction of the RS framework, following from the composite nature of
the top quark, we conclude that the corrections to $\AFBt$ are
necessarily negative. This feature is illustrated in
\Fig{fig:RSpredictions}, which shows the predictions for the absolute
RS corrections to the forward-backward asymmetry in the $p \bar p$
frame as a function of $c_{t_L}$ and $c_{t_R}$. The figures have been
obtained including only the KK-gluon corrections to $\tilde C_{u \bar
  u}^{V,A}$ and employing $\Mkk = 1 \, {\rm TeV}$, $c_{u_L} =c_{d_L} =
-0.63$, $c_{u_R} = -0.68$, $c_{d_R} = -0.66$, $c_{c_L} =c_{s_L} =
-0.56$, and $c_{c_R} = -0.53$, $c_{s_R} = -0.63$, as well as setting
all minors of $Y_{u,d}$ equal (only $Y_t = (Y_u)_{33}$ is allowed to
vary in order to reproduce the observed top-quark mass). Both panels
show clearly that in the whole $c_{t_L}$--$c_{t_R}$ plane the
corrections to $(\AFBt)_{\rm RS}^{p \bar p}$ interfere destructively
with the SM. However, even for an optimistic value of $\Mkk = 1 \,
{\rm TeV}$, corresponding to a mass of the lightest KK gluon of around
$2.5 \, {\rm TeV}$, we find that after imposing the $Z \to b \bar b$
constraints\footnote{For a detailed discussion see
  \cite{Casagrande:2010si}.} the maximal attainable effects amount to
not even $-0.05\%$ ($-0.10\%$) in the minimal (extended) RS model
based on an $SU(2)_L \times U(1)_Y$ ($SU(2)_R \times SU(2)_L \times
U(1)_X \times P_{LR}$) bulk gauge group.\footnote{Including all RS
  corrections, we find that for the three parameter points considered
  before, the $t \bar t$ forward-backward asymmetry is shifted by
  $-0.04\%$, $-0.05\%$, and $-0.05\%$ with respect to the SM value.}
The parameter regions compatible with the $Z \to b \bar b$ data are
colored green in \Fig{fig:RSpredictions}. While this constraint is
very stringent in the minimal model, restricting the allowed parameter
space to a thin stripe with $c_{t_L} \in [-0.60, -0.49]$, it does not
pose a tight bound in the case of the extended scenario allowing for
$c_{t_L} \in [-0.60, 1]$. Since the KK-gluon corrections decouple as
$1/\Mkk^2$, employing $\Mkk = 2 \, {\rm TeV}$ instead of $1 \, {\rm
  TeV}$ implies that the possible effects in the $t \bar t$
forward-backward asymmetry are less than $-0.03\%$.

Our results should be contrasted with the analysis
\cite{Djouadi:2009nb}, which finds positive corrections to the $t
\bar t$ forward-backward asymmetry of up to $5.6\%$ ($7\%$) arising
from KK gluons ($Z^\prime$-boson exchange) at LO. In the latter
article, sizable corrections to $\tilde C_{ q \bar q}^A$ arise since
the left- and right-handed components of the light-quark fields are
localized at different ends of the extra dimension by choosing
$c_{u_L} = c_{d_L} \in [ -0.4, 0.4]$ (IR-localized) and $c_{u_R} =
c_{d_R} = -0.8$ (UV-localized).\footnote{Notice that the convention of
  the bulk mass parameters used in \cite{Djouadi:2009nb} differs from
  ours by an overall sign.} In an anarchic approach to flavor, such a
choice is in conflict with observation, because it fails to reproduce
the hierarchies of light-quark masses and mixings.

\section{Conclusions and Outlook}
\label{sec:concl}

In this work we have studied the interplay between new-physics LO and
NLO effects to the top-quark forward-backward asymmetry $\AFBt$ within
RS models. In scenarios with flavor anarchy, the dominant
contributions to $t\bar t$ production arise from $s$-channel exchange
of KK gluons. The axial-vector couplings to light quarks are
suppressed due to both their UV localization as well as the close
separation of wave functions of different chiralities in the extra
dimension. The resulting exponential depletion inhibits potentially
large $s$-channel contributions to the asymmetry at tree level. The
suppression of flavor-changing $t\bar u$ couplings turns out to be
even stronger, such that it is impossible to obtain sizable effects in
the $t$ channel as well. Consequently, if the quarks are localized in
the extra dimension such that their masses and flavor mixings are
correctly reproduced, LO corrections to the forward-backward asymmetry
in RS models are deemed to be far too small to be able to explain the
observed discrepancy between experiment and SM expectation in $\AFBt$.

\begin{figure}[t!]
\begin{center}
\includegraphics[width=7.5cm]{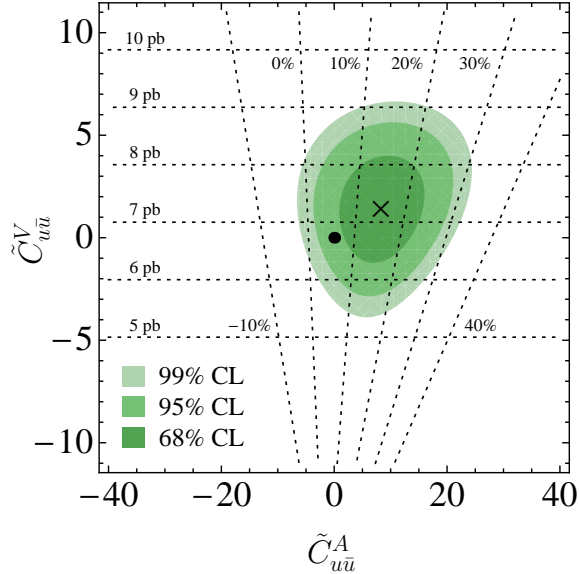}
\end{center}
\vspace{-7.5mm}
\begin{center}
  \parbox{15.5cm}{\caption{\label{fig:cvca} Results of a combined fit
      to $\sigtot$, the last bin of $\dsig$, and the value of
      $(\AFBt)^{p \bar p}$ allowing for new physics in $s$-channel
      exchange. The green contours indicate, from dark to light, the
      experimentally favored regions of 68\%, 95\%, and 99\%
      probability in the $\tilde C_{u \bar u}^A$--$\tilde C_{u \bar
        u}^V$ plane. The horizontal (almost vertical) dashed lines
      correspond to the value of the total $t \bar t$ cross section
      (forward-backward asymmetry in the $p \bar p$ frame). Further
      details can be found in the text.}}
\end{center}
\end{figure}

We have furthermore shown that vector currents resulting from KK-gluon
exchange yield a positive contribution to the asymmetric cross section
$\sigma_a$ at NLO and are not subject to any suppression related to
the localization of quark wave functions. Numerically, we found that
the ratio of the products of vector and axial-vector couplings of the
light quarks and the top quark generically satisfies $(g_V^q
g_V^t)/(g_A^q g_A^t ) = {\cal O} (10^3)$. This strong enhancement
implies that, despite their loop suppression, NLO vector-current
contributions to the asymmetry exceed the LO axial-vector correction
by typically a factor of around $100$. However, tree-level vector
currents, arising from KK-gluon exchange, tend to also enhance the
symmetric cross section $\sigma_s = \sigtot$, which enters the
normalization of the $t \bar t$ forward-backward asymmetry. Our
numerical analysis reveals that in practice the NLO vector contribution
to $\sigma_a$ is cancelled in large parts by the LO contribution to
$\sigma_s$, so that the resulting sensitivity of $\AFBt =
\sigma_a/\sigma_s$ to vector currents is not very pronounced. This
feature not only limits the magnitude of the possible RS contributions
in $\AFBt$ to far below the percent level, but also leads to the
robust prediction, related to the compositeness of the top quark, that
these corrections are necessarily destructive.  These findings are
largely model-independent, as they do not depend strongly on the exact
realization of the electroweak bulk gauge group, the choice of fermion
representations, or the precise nature of the Higgs sector of the
considered RS setup.

As our arguments are sufficiently general, they do not only apply to
the RS framework but to the broader class of models with new heavy
vector states that have small axial-vector couplings to light
quarks. In particular, many new-physics models which address the
flavor problem via a Froggatt-Nielsen-type mechanism belong to the
latter category. The aforementioned cancellation of vector
contributions between the numerator and denominator of $\AFBt$
suggests that in such new-physics scenarios, irrespectively of their
sign, large contributions to the $t \bar t$ forward-backward asymmetry
are essentially impossible to achieve, once the experimentally
available information on $\sigtot$ and the high-energy tail of the $t
\bar t$ invariant mass spectrum $\dsig$ is taken into account. This is
illustrated in Figure~\ref{fig:cvca}, which shows the results of a
global fit to the available $t \bar t$ data (see (\ref{eq:AFBexp}) and
(\ref{eq:EXPss})) in the presence of new physics in the $s$
channel. The colored contours indicate the experimentally preferred
region in the $\tilde C_{u \bar u}^{A}$--$\tilde C_{u \bar u}^{V}$
plane. From the shape and location of the favored area, one infers
that a non-zero vector coefficient $\tilde C_{u \bar u}^V$ alone does
not lead to a significant improvement in the quality of the fit, but
that large corrections to the axial-vector coefficient $\tilde C_{u
  \bar u}^A$ are needed to get from the SM point (black dot) at
$(0,0)$ to the best-fit value (black cross) at $(8.3, 1.4)$. In fact,
requiring the three $t \bar t$ predictions to be within the global
95\% (99\%) CL region allows for maximal values $(\AFBt)^{p \bar p}$
of $5.8\%$ ($6.0\%$) from vector contributions alone. The
corresponding point in the $\tilde C_{u \bar u}^{A}$--$\tilde C_{u
  \bar u}^{V}$ plane is located at $(0,-1.8)$ $\big ( (0,-3.1) \big
)$. We conclude from these general observations that a large $t \bar
t$ forward-backward asymmetry inevitably has to arise from tree-level
effects involving either axial-vector currents in the $s$ channel with
flavor-specific couplings of opposite sign to light quarks and top
quarks or large flavor-changing currents in the $t$ channel. However,
both options are difficult to realize in any explicit construction of
physics beyond the SM without invoking additional {\it ad hoc}
assumptions about the flavor structure of the light-quark
sector. There thus seems to be a generic tension between having large
effects in $\AFBt$ and achieving a natural solution to the flavor
problem.

\subsubsection*{Acknowledgments}

We are grateful to V.~Ahrens, A.~Ferroglia, M.~Neubert, B.~Pecjak, and
L.~Yang for useful discussions. The Feynman diagrams shown in this
work are drawn using {\tt FeynArts} \cite{Hahn:2000kx}.  This
research is supported in part by the German Federal Ministry for
Education and Research grant 05H09UME (``Precision Calculations for
Collider and Flavour Physics at the LHC''), the Helmholtz-Institut
Mainz, and the Research Centre ``Elementary Forces and Mathematical
Foundations'' funded by the Excellence Initiative of the State of
Rhineland-Palatinate. U.H. would like to thank the Aspen Center for
Physics, where part of this research was performed.

\begin{appendix}

\section{Higgs-Boson Phase-Space Factors}
\label{app:phasespace}

\renewcommand{\theequation}{A\arabic{equation}}
\setcounter{equation}{0}

In this appendix we present the explicit form of the phase-space
factors appearing in the Higgs-boson contribution to the
charge-symmetric and -asymmetric part of the $t \bar t$ cross
section. The functions $f_{S,A} (z)$ introduced in (\ref{eq:SLONP})
and (\ref{eq:ALONP}) read
\beq \label{eq:fSfA}
\begin{split}
  f_S (z) & = -\frac{\beta \rho}{72} \left[ 1 +\frac{\rho \left ( 1-z
      \right )}{2}+ \frac{\rho \left(4+ \rho \left (1-z \right )^2
      \right) }{8 \beta } \, \ln \left(\frac{ 2 \left ( 1+ \beta
        \right ) - \rho \left (1 - z\right )}{2 \left ( 1- \beta
        \right ) -\rho \left ( 1 - z
        \right )}\right) \right] \,, \\
  f_A (z) & = \frac{\rho }{144} \left[ 1-\rho + \frac{\rho
      \left(4+ \rho \left (1-z \right )^2 \right)}{4 } \, \ln
    \left(\frac{\rho \left(4 z + \rho \left (1-z \right )^2
        \right)}{(2 - \rho \left (1-z \right ) )^2}\right) \right ]\,,
\end{split}
\eeq 
where $z = m_h^2/m_t^2$, $\beta = \sqrt{1-\rho}$, and $\rho = 4
m_t^2/\hat s$.

\section{Wilson Coefficients in the ZMA}
\label{app:ZMA}

\renewcommand{\theequation}{B\arabic{equation}}
\setcounter{equation}{0}

In the following, we present the ZMA results for the Wilson
coefficients in \eq{eq:wilsonexplicit}. Specializing to the case of
the up quark ($q=u$), we find
\begin{align} \label{eq:wilsonuuZMA}
  C^{(V, 8)}_{u \bar u, \parallel} & = -\frac{4 \pi \alpha_s}{\Mkk^2}
  \Bigg [ \frac{1}{2L} -\frac{F^2(c_{t_R}) \left ( 2 c_{t_R} + 5
    \right )}{4 (2 c_{t_R} +3)^2} -\frac{F^2(c_{t_L}) \left ( 2
      c_{t_L} + 5 \right )}{4 (2 c_{t_L} +3)^2} \nonumber \\ & \phantom{xx}
  -\frac{F^2(c_{u_R})}{4\left |(M_u)_{11} \right |^2} \sum_{i=1,2,3}
  \frac{(2 c_{u_i} + 5) \left |(M_u)_{1i} \right |^2 }{(2 c_{u_i}
    +3)^2} -\frac{F^2(c_{u_L})}{4 \left |(M_u)_{11} \right |^2}
  \sum_{i=1,2,3} \frac{(2 c_{Q_i} + 5) \left |(M_u)_{i1} \right |^2
  }{(2 c_{Q_i} +3)^2} \nonumber \\ &\phantom{xx} + \frac{L}{2} \,
  \frac{F^2(c_{t_R}) \hspace{0.25mm} F^2(c_{u_R})}{(2 c_{t_R}+3) \left
      |(M_u)_{11} \right |^2} \sum_{i=1,2,3} \frac{(c_{u_i} + c_{t_R}
    +3) \left |(M_u)_{1i} \right |^2 }{(2 c_{u_i} +3)
    (c_{u_i} + c_{t_R} +2)} \nonumber \\
  &\phantom{xx} + \frac{L}{2} \, \frac{F^2(c_{t_L}) \hspace{0.25mm}
    F^2(c_{u_L})}{(2 c_{t_L}+3) \left |(M_u)_{11} \right |^2}
  \sum_{i=1,2,3} \frac{(c_{Q_i} + c_{t_L} +3) \left |(M_u)_{i1} \right
    |^2 }{(2 c_{Q_i} +3)
    (c_{Q_i} + c_{t_L} +2)}\Bigg ] \,, \nonumber \\[6mm]
  C^{(V, 8)}_{u \bar u, \perp} & = -\frac{4 \pi \alpha_s}{\Mkk^2}
  \Bigg[ \frac{1}{2L} -\frac{F^2(c_{t_R}) \left ( 2 c_{t_R} + 5 \right
    )}{4 (2 c_{t_R} +3)^2} -\frac{F^2(c_{t_L}) \left ( 2 c_{t_L} + 5
    \right )}{4 (2 c_{t_L} +3)^2} \nonumber \\ & \phantom{xx}
  -\frac{F^2(c_{u_R})}{4\left |(M_u)_{11} \right |^2} \sum_{i=1,2,3}
  \frac{(2 c_{u_i} + 5) \left |(M_u)_{1i} \right |^2 }{(2 c_{u_i}
    +3)^2} -\frac{F^2(c_{u_L})}{4 \left |(M_u)_{11} \right |^2}
  \sum_{i=1,2,3} \frac{(2 c_{Q_i} + 5) \left |(M_u)_{i1} \right |^2
  }{(2 c_{Q_i} + 3)^2} \nonumber \\ &\phantom{xx} + \frac{L}{2} \,
  \frac{F^2(c_{t_L}) \hspace{0.25mm} F^2(c_{u_R})}{(2 c_{t_L}+3) \left
      |(M_u)_{11} \right |^2} \sum_{i=1,2,3} \frac{(c_{u_i} + c_{t_L}
    +3) \left |(M_u)_{1i} \right |^2 }{(2 c_{u_i} +3)
    (c_{u_i} + c_{t_L} +2)} \nonumber \\
  &\phantom{xx} + \frac{L}{2} \, \frac{F^2(c_{t_R}) \hspace{0.25mm}
    F^2(c_{u_L})}{(2 c_{t_R}+3) \left |(M_u)_{11} \right |^2}
  \sum_{i=1,2,3} \frac{(c_{Q_i} + c_{t_R} +3) \left |(M_u)_{i1} \right
    |^2 }{(2 c_{Q_i} +3)(c_{Q_i} + c_{t_R} +2)} \Bigg] \,,
\end{align}
and similar relations hold in the case of the remaining light quarks
$q = d, s, c$. Here $(M_u)_{ij}$ are the minors of the up-type Yukawa
matrix $Y_u$. For the $t$-channel Wilson coefficients in the vector
channel, we obtain
\bea \label{eq:wilsontuZMA}
\begin{split}
  C^{(V, 8)}_{t \bar u, \parallel} & = -\frac{\pi \alpha_s}{\Mkk^2} \,
  L \Bigg [ \frac{F^2(c_{t_R}) \hspace{0.25mm} F^2(c_{u_R})\left
      |(M_u)_{13} \right |^2 }{ ( 2 c_{t_R} + 3 ) ( c_{t_R} + 1) \left
      |(M_u)_{11} \right |^2} +\frac{F^2(c_{t_L}) \hspace{0.25mm}
    F^2(c_{u_L}) \left |(M_u)_{31} \right |^2 }{(2 c_{t_L} + 3)
    (c_{t_L}+1) \left |(M_u)_{11} \right |^2} \Bigg] \,,
  \\[1mm]
  C^{(V, 1)}_{t \bar u, \parallel} & = -\frac{\pi \alpha_e}{\Mkk^2} \,
  \frac{L}{s_w^2 c_w^2} \, \Bigg [ (T_3^u-Q_u s_w^2)^2 \,
  \frac{F^2(c_{t_L}) \hspace{0.25mm} F^2(c_{u_L}) \left |(M_u)_{31}
    \right |^2 }{(2 c_{t_L} + 3) (c_{t_L}+1) \left |(M_u)_{11} \right
    |^2} \\ & \hspace{2.75cm} + \left ( s_w^2 Q_u \right )^2
  \frac{F^2(c_{t_R}) \hspace{0.25mm} F^2(c_{u_R}) \left |(M_u)_{13}
    \right |^2 }{( 2 c_{t_R} + 3 ) ( c_{t_R} + 1) \left |(M_u)_{11}
    \right |^2} \Bigg ] \\ & \phantom{xx} - \frac{\pi \alpha_e
    Q_u^2}{\Mkk^2} \, L \Bigg [ \frac{F^2(c_{t_R}) \hspace{0.25mm}
    F^2(c_{u_R})\left |(M_u)_{13} \right |^2 }{ ( 2 c_{t_R} + 3 ) (
    c_{t_R} + 1) \left |(M_u)_{11} \right |^2} +\frac{F^2(c_{t_L})
    \hspace{0.25mm} F^2(c_{u_L}) \left |(M_u)_{31} \right |^2 }{ (2
    c_{t_L} + 3) (c_{t_L}+1) \left |(M_u)_{11} \right |^2} \Bigg ] \,.
\end{split}
\eea
For completeness we also give the result of the Higgs-boson
contribution to the $t$ channel. We find for the dimensionless
coefficient
\begin{equation}
\tilde C^{S}_{t \bar u}  = 
    \left | (g_h^u)_{13} \right |^2 + \left | (g_h^u)_{31} \right |^2 \,,
\end{equation}
where
\begin{align}
\begin{split}
(g_h^u)_{13}=
&\,-\frac{m_u}v \frac{m_t^2}{\Mkk^2} \left(W_u^\dagger\,\,
   \mbox{diag}\left[ \frac{1}{1-2c_{u_i}} 
   \left( \frac{1}{F^2(c_{u_i})} 
   - 1 + \frac{F^2(c_{u_i})}{3+2c_{u_i}} \right) \right] 
   W_u\right)_{13} \\
&\, +\frac {v^2 }{3\sqrt{2} \,  M_{\rm
       KK}^2} \; \left(U_u^\dagger \; {\rm diag} \left[ F(c_{Q_i})
   \right] \hspace{0.25mm} Y_u \hspace{0.25mm} Y_u^\dagger
   \hspace{0.25mm} Y_u \; {\rm diag } \left[ F(c_{u_i})
   \right] W_u\right)_{13}\,, \\   
(g_h^u)_{31}= &\,-\frac{m_u}v \frac{m_t^2}{\Mkk^2} \left(U_u^\dagger\,\,
   \mbox{diag}\left[ \frac{1}{1-2c_{Q_i}} 
   \left( \frac{1}{F^2(c_{Q_i})} 
   - 1 + \frac{F^2(c_{Q_i})}{3+2c_{Q_i}} \right) \right] 
   U_u\right)_{31} \\
&\, +\frac {v^2 }{3 \sqrt{2} \, M_{\rm
       KK}^2} \; \left(U_u^\dagger \; {\rm diag} \left[ F(c_{Q_i})
   \right] \hspace{0.25mm} Y_u \hspace{0.25mm} Y_u^\dagger
   \hspace{0.25mm} Y_u \; {\rm diag } \left[ F(c_{u_i})
   \right] W_u\right)_{31}\,, \\       
\end{split}
\end{align}
and the $3\times 3$ unitary matrices $U_u$ and $W_u$ are defined
through
\begin{equation}
\displaystyle
{\rm diag }\left [F(c_{Q_i}) \right ]\, Y_u\, {\rm diag } \left [F(c_{u_i}) \right ]=
\frac{\sqrt 2}v\;U_u\,{\rm diag }[m_u,m_c,m_t]\,W_u^\dagger\,.
\end{equation}

\section{RG Evolution of  the Wilson Coefficients}
\label{app:RGE}

\renewcommand{\theequation}{C\arabic{equation}}
\setcounter{equation}{0}

This appendix contains analytic formulas relating the Wilson
coefficients evaluated at the top-quark mass scale $m_t$ with their
initial conditions calculated at $\Mkk \gg m_t$. Since in the RS model
the $t$-channel Wilson coefficients $\tilde C_{t\bar u}^{V}$ and
$\tilde C_{t \bar u}^{S}$ turn out to be numerically irrelevant, we
will not consider their running in the following. We perform the RG
evolution at leading-logarithmic accuracy, \ie, at one-loop order,
neglecting tiny effects that arise from the mixing with QCD penguin
operators. For the $s$-channel Wilson coefficients entering the
formulas (\ref{eq:EFTss}) and (\ref{eq:AFBEFT}), we find for $P = V,
A$, 
\beq \label{eq:c1}
\tilde C_{q\bar q}^P(m_t) = \left ( \frac{2}{3 \eta^{4/7}} +
  \frac{\eta^{2/7}}{3} \right ) \tilde C_{q\bar q}^P(\Mkk) \,,
\eeq 
where $\eta \equiv \alpha_s (\Mkk)/\alpha_s(m_t)$ is the ratio of
strong coupling constants evaluated at the relevant scales $\Mkk$ and
$m_t$. 

In order to get an idea of the potential impact of RG effects, we
evaluate (\ref{eq:c1}) using $\alpha_s(M_Z)=0.139$, $\Mkk=1\TeV$, and
$m_t=173.1\GeV$, which leads to $\eta=0.803$ at one-loop order. We
obtain
\beq \label{eq:c2}
\tilde C_{q\bar q}^P(m_t) = 1.07\,\tilde C_{q\bar q}^P(\Mkk)\,,
\eeq
from which we conclude that the RG evolution increases the Wilson
coefficients $\tilde C_{q\bar q}^P$ by about $7\%$ with respect to the
values quoted in \Tab{tab:WC}. Operator mixing thus represents only a
numerically subdominant effect.

\section{Parameter Points}
\label{app:points}

\renewcommand{\theequation}{D\arabic{equation}}
\setcounter{equation}{0}

In order to make our work self-contained, we specify in this appendix
the complete set of model parameters, namely the bulk mass parameters
of the quark fields and the Yukawa matrices, corresponding to the
three parameter points used in our numerical analysis. All the
parameter sets given below have been obtained by random choice,
subject to the constraints that the absolute value of each entry in
$Y_{u,d}$ is between $1/3$ and $3$, and that the Wolfenstein
parameters $\bar\rho$ and $\bar\eta$ agree with experiment within
errors. The bulk mass parameters have then been determined using the
warped-space Froggatt-Nielsen formulas given in
\cite{Casagrande:2008hr}, which guarantees that the quark masses and
mixings are correctly reproduced. For further details concerning the
algorithm used to scan the parameter space of the RS model, the
interested reader is referred to \cite{Bauer:2009cf}.

Our first parameter point is specified by the following bulk mass
parameters\footnote{Here and below, results are given to at least
  three significant digits.}
\beq\label{eq:cparameter1}
\begin{aligned}
   c_{Q_1} &= -0.611 \,, \qquad
    c_{Q_2} &= -0.580 \,, \qquad
    c_{Q_3} &= -0.407 \,, \\
   c_{u_1} &= -0.688 \,, \qquad
    c_{u_2} &= -0.550 \,, \qquad
    c_{u_3} &= +0.091 \,, \\
   c_{d_1} &= -0.665 \,, \qquad
    c_{d_2} &= -0.627 \,, \qquad
    c_{d_3} &= -0.577 \,,
\end{aligned}
\eeq
and Yukawa matrices 
\beq\label{eq:yukawas1}
\begin{split}
  \bm{Y}_u &= \left(
    \begin{array}{rrr} 
      -1.303-0.364 \hspace{0.5mm} i &
      ~~-1.215+0.089 \hspace{0.5mm} i~~ &        
      -1.121-1.679 \hspace{0.5mm} i \\
      1.857+1.199 \hspace{0.5mm} i &       
      ~~2.038+1.105 \hspace{0.5mm} i~~ &       
      -0.484-0.193 \hspace{0.5mm} i \\
      -1.052+0.546 \hspace{0.5mm} i & 
      ~~-2.833+0.191 \hspace{0.5mm}
      i~~ & -1.287-1.141 \hspace{0.5mm} i
   \end{array}
 \right) , \\
 \bm{Y}_d &= \left(
   \begin{array}{rrr}
     -0.661-1.118 \hspace{0.5mm} i &       
     ~~-0.075-0.656 \hspace{0.5mm} i~~ &       
     ~~0.141-0.465 \hspace{0.5mm} i \\
     -2.070+1.364 \hspace{0.5mm} i &      
     ~-2.518+1.435 \hspace{0.5mm} i~~ &       
     0.717-0.165 \hspace{0.5mm} i \\
     0.306+2.830 \hspace{0.5mm} i &       
     ~~0.034-0.350 \hspace{0.5mm} i~~ &       
     -0.951-0.829 \hspace{0.5mm} i
   \end{array}
   \right) .
\end{split}
\eeq
The second parameter point is given by 
\beq\label{eq:cparameter2}
\begin{aligned}
   c_{Q_1} &= -0.646 \,, \qquad
    c_{Q_2} &= -0.573 \,, \qquad
    c_{Q_3} &= -0.449 \,, \\
   c_{u_1} &= -0.658 \,, \qquad
    c_{u_2} &= -0.513 \,, \qquad
    c_{u_3} &= +0.480 \,, \\
   c_{d_1} &= -0.645 \,, \qquad
    c_{d_2} &= -0.626 \,, \qquad
    c_{d_3} &= -0.578 \,,
\end{aligned}
\eeq
and
\beq\label{eq:yukawas2}
\begin{split}
 \bm{Y}_u &= \left(
   \begin{array}{rrr}
     0.637-1.800 \hspace{0.5mm} i &
     ~~1.518-2.209 \hspace{0.5mm} i~~ &       
     0.904+0.146 \hspace{0.5mm} i \\
     0.219-0.207 \hspace{0.5mm} i &      
     ~~-0.333-0.942 \hspace{0.5mm} i~~ &     
     0.597+0.020 \hspace{0.5mm} i \\
     1.829+1.538 \hspace{0.5mm} i & 
     ~~-0.018+1.772 \hspace{0.5mm} i~~ &
     -1.258+1.265 \hspace{0.5mm} i
  \end{array}
\right) , \\
\bm{Y}_d &= \left(
  \begin{array}{rrr}
    -2.835-0.946 \hspace{0.5mm} i &      
    ~~-0.404+0.746 \hspace{0.5mm} i~~ &      
    -1.135+0.060 \hspace{0.5mm} i \\
    0.724-0.350 \hspace{0.5mm} i &      
    ~-2.214-0.555 \hspace{0.5mm} i~~ &      
    0.610-0.051 \hspace{0.5mm} i \\
    0.701-0.101 \hspace{0.5mm} i &      
    ~~-0.154+0.104 \hspace{0.5mm} i~~ &      
    1.514+0.919 \hspace{0.5mm} i
  \end{array}
  \right) .
\end{split}
\eeq
Finally, our third parameter point features 
\beq\label{eq:cparameter3}
\begin{aligned}
   c_{Q_1} &= -0.624 \,, \qquad
    c_{Q_2} &= -0.563 \,, \qquad
    c_{Q_3} &= -0.468 \,, \\
   c_{u_1} &= -0.712 \,, \qquad
    c_{u_2} &= -0.560 \,, \qquad
    c_{u_3} &= +0.899 \,, \\
   c_{d_1} &= -0.659 \,, \qquad
    c_{d_2} &= -0.642 \,, \qquad
    c_{d_3} &= -0.571 \,,
\end{aligned}
\eeq
and 
\beq\label{eq:yukawas3}
\begin{split}
 \bm{Y}_u &= \left(
  \begin{array}{rrr}        
    -0.541+1.517 \hspace{0.5mm} i &     
    ~~-1.083+1.857 \hspace{0.5mm} i~~ &       
    1.718-2.057 \hspace{0.5mm} i \\
    0.359-1.713 \hspace{0.5mm} i &      
    ~~-2.208+1.404 \hspace{0.5mm} i~~ &      
    -1.160+0.886 \hspace{0.5mm} i \\
    -1.172-0.543 \hspace{0.5mm} i &      
    ~~-0.116-0.238 \hspace{0.5mm} i~~ &       
    -0.669-1.688 \hspace{0.5mm} i
  \end{array}
\right) , \\
\bm{Y}_d &= \left(
  \begin{array}{rrr}
    -0.878-1.677\hspace{0.5mm} i &      
    ~~0.190+0.573 \hspace{0.5mm} i~~ &      
    -0.817+2.663 \hspace{0.5mm} i \\
    -1.792+0.861 \hspace{0.5mm} i &      
    ~-2.880+0.132 \hspace{0.5mm} i~~ &      
    -0.070-1.151 \hspace{0.5mm} i \\
    -1.679+1.588 \hspace{0.5mm} i &     
    ~~0.972+0.615 \hspace{0.5mm} i~~ &      
    1.421+0.981 \hspace{0.5mm} i
  \end{array}
  \right) .
\end{split}
\eeq
Here $c_{Q_1} = c_{u_L} = c_{d_L}$, $c_{u_1} = c_{u_R}$, and $c_{d_1}
= c_{d_R}$ and similarly in the case of the second and third quark
generation.

\end{appendix}

\end{document}